\documentclass[twocolumn,reprint,amsmath,amssymb,showpacs,prb,aps,superscriptaddress]{revtex4-1}
\usepackage{graphicx}
\usepackage{times}
\usepackage{txfonts}
\usepackage{setspace}
\usepackage{CJK}
\usepackage[colorlinks=true,citecolor=blue,urlcolor=blue]{hyperref}

\newcommand{\ii}{{\rm i}}
\def\tr{ {\rm tr}}

\begin{document}
%\newfloat
\newcommand{\eq}[1]{Eq.~(\ref{#1})}
\newcommand{\fig}[1]{Fig.~\ref{#1}}
\newcommand{\App}[1]{Appendix ~\ref{#1}}

\begin{CJK*}{UTF8}{}
\title{Quantum Impurity in Luttinger Liquid: Universal Conductance with Entanglement Renormalization}
\CJKfamily{bsmi}

\author{Ya-Lin Lo (羅雅琳)}
%\author{Ya-Lin Lo}
\affiliation{ Department of Physics, National Taiwan University, Taipei 10607, Taiwan}
\affiliation{ Center of Theoretical Sciences, National Taiwan University, Taipei 10607, Taiwan}

\author{Yun-Da Hsieh  (謝昀達)}
%\author{Yun-Da Hsieh }
\affiliation{ Department of Physics, National Taiwan University, Taipei 10607, Taiwan}
\affiliation{ Center of Theoretical Sciences, National Taiwan University, Taipei 10607, Taiwan}

\author{ Chang-Yu Hou }
\affiliation{Department of Physics, California Institute of Technology, Pasadena, CA}

\author{Pochung Chen (陳柏中)}
%\author{Pochung Chen}
\email{pcchen@phys.nthu.edu.tw}
\affiliation{ Department of Physics, National Tsing Hua University, Hsinchu 30013, Taiwan}
\affiliation{ Physics Division, National Center for Theoretical Sciences, Hsinchu 30013, Taiwan}

\author{Ying-Jer Kao (高英哲)}
%\author{Ying-Jer Kao }
\email{yjkao@phys.ntu.edu.tw}
\affiliation{ Department of Physics, National Taiwan University, Taipei 10607, Taiwan}
\affiliation{ Center of Advanced Study in Theoretical Science, National Taiwan University, Taipei 10607, Taiwan}
\affiliation{ Physics Division, National Center for Theoretical Sciences, Hsinchu 30013, Taiwan}
%\affiliation{ Center for Quantum Science and Engineering, National Taiwan University, Taipei 106, Taiwan}
\date{\today}

\begin{abstract}
We study numerically the universal conductance of Luttinger liquids wire with a single impurity via the Muti-scale Entanglement Renormalization Ansatz (MERA). 
The scale invariant MERA  provides an efficient way to extract scaling operators and scaling dimensions for both the bulk and the boundary conformal field theories. 
By utilizing the key relationship between the conductance tensor and ground-state correlation function, the universal conductance can be evaluated within the framework of the boundary MERA.
We construct the boundary MERA to compute the correlation functions and scaling dimensions for the Kane-Fisher fixed points by modeling the single impurity as a junction (weak link) of two interacting wires. 
We show that the universal behavior of the junction can be easily identified within the MERA
and argue that the boundary MERA framework has tremendous potential to classify the fixed points in general multi-wire junctions.
\end{abstract}

\pacs{05.10.Cc, %Renormalization group methods
02.70.-c, %Computational techniques; simulations
05.60.Gg, % Quantum transport
05.50.+q %Lattice theory and statistics
}
\maketitle
\end{CJK*}

\section{Introduction}
Recent advances in nano-fabrication allow device miniaturization to the molecular scale. Devices such as single molecule junctions connecting to multiple metallic leads are promising candidates as the building blocks for molecular electronics.\cite{Joachim:2005fk,dnameasure:2012} Furthermore, it is now possible to confine electrons in one-dimensional (1D) quantum wires, where Luttinger liquid (LL) can be realized with short ranged electron-electron interactions.~\cite{Laroche:2014qf,carbontube:2003,carbonnanotube:nature1999,carbontube1999,PhysRevLett.99.036802,PhysRevB.62.R10653} As a result, fabrications of junctions of multiple LL wires are within the reach of current experimental technology. Therefore, understanding properties of the multi-wire junction, such as the linear conductance, are of current interest.

Theoretically, one-dimensional (1D) interacting quantum systems enjoy a special status as there exists a plethora of analytical and numerical methods. In particular, for 1D critical systems, we can use powerful theoretical tools such as the conformal field theory (CFT) and the renormalization group (RG) to analyze the physical properties.~\cite{Cardy2010,Affleck2010} For instance, the presence of a potential barrier (impurity) leads to a boundary RG fixed point that determines the transport of a 1D interacting LL.~\cite{Luttinger63,Kane1992,Kane1992b,Furusaki1993} The CFT description suggests that a conformally invariant boundary condition (CIBC) will be associated with a boundary RG fixed point due to the presence of the impurity.~\cite{Wong1994} A complemental RG approach with fermionic description instead of the standard bosonization procedure can also be used at weak interaction and provides a route to capture the non-Luttinger liquid behaviors in 1D quantum wires.~\cite{Matveev1993} These analytical approaches have yielded great success in studying various 1D quantum impurity problems, such as Kondo impurities,~\cite{Affleck1991} resonance tunnelings~\cite{Nayak1999} and junctions of quantum wires.~\cite{Oshikawa2006}

On the other hand, numerical studies on the LLs with impurities have provided useful insights into the properties of the RG fixed points,~\cite{Andergassen2004,Hamamoto2008,Freyn2011} and have aided the identification of new fixed points for more complicated structures.~\cite{PhysRevB.71.205327,PhysRevLett.94.136405,Rahmani2010} However, it is difficult to simulate 1D critical systems, of which the LL is an example, because reaching scale invariance in order to  capture the true  power law correlations requires large system sizes. A recent proposal based on  tensor network states called the multi-scale entanglement renormalization ansatz (MERA) has been shown to overcome these difficulties in  simulating scale invariant critical systems.~\cite{ER} The key concept of the MERA is to keep only the long-range entanglement of the system during the real-space RG transformation. In particular, MERA in its scale invariant form allows one to extract the universal properties such as  critical exponents,  scaling dimensions and  long-range power law correlations. Moreover, since the effects of an impurity can be included by introducing an impurity defined boundary,  the boundary MERA is able to capture the boundary RG fixed points and serves as an ideal tool to study quantum impurity problems in 1D quantum critical systems.~\cite{MERABCFT}

%{\bf use MERA to study RG flow and fixed point}

With the density-matrix-renormalization-group (DMRG)  as the primary numerical scheme currently to study quasi-1D interacting systems,~\cite{White,Schollwock:2005vn} it is worthwhile to discuss briefly how and where the boundary MERA scheme can have advantage over DMRG. First, since the finite-size DMRG calculation rarely reaches scale invariance, it becomes non-trivial to extract properties of boundary RG fixed points due to the presence of an impurity in a 1D critical system. Often, a finite size scaling or further manipulation on the numerical data is required to extract the necessary information in order to show the effects of the boundary.~\cite{Meden2003} Specifically, previous attempts using DMRG to obtain the fixed point universal conductance of a multi-wire junction has its limitations: it is necessary to perform a conformal transformation of the correlation functions to map the semi-infinite wire system to a finite strip, and a second boundary term has to be added to cap the system in order to perform a finite-size DMRG.~\cite{Rahmani2010,Rahmani2012} The mapping between the two boundary Hamiltonians is obtained exactly in the non-interacting case, and is argued to remain valid in the interacting case.~\cite{Rahmani2010} Even with this manipulation, it is still necessary to perform calculations in a large enough system size to reach the scaling invariant at the middle of the wire. However, it is not straightforward to know \textit{a priori} how large the system size has to be to obtain the scale invariant properties of RG fixed points, especially for unknown RG fixed points. On the other hand, while an infinite DMRG calculation can reach the scale invariance limit and displays the power law correlations,~\cite{Karrasch2012} it requires translational invariance. Addition of an impurity into such a calculation can be numerically costly as the translational invariance is broken  explicitly. A numerical method that can explicitly preserve scale invariance in the presence of an impurity, and perform direct simulations on the (semi-)infinite chains is coveted. %, and the boundary MERA framework is a promising candidate.

In this paper, we employ the boundary MERA to study the simplest 1D quantum transport with an impurity: a single weak link (potential barrier) in a spinless LL. As shown by Kane and Fisher,~\cite{Kane1992} there exists two possible RG fixed points:  a  total reflection fixed point with two disconnected wires when the electron-electron interaction in the lead is repulsive, and a perfect transmission fixed point when the interaction is attractive. Although numerical analysis based on DMRG and functional RG shows evidences in support of these conclusions,~\cite{PhysRevB.46.10866,PhysRevB.54.R9643,Enss2005,Andergassen2004,Andergassen2006} a \textit{direct} computation of  correlation functions on the semi-infinite wires with a junction remains illusive. Using a MERA that explicitly preserves  the scale invariance, we are able to compute the current-current correlation functions, spin-spin correlation functions, and the scaling dimensions of a 1D LL in the presence of an impurity. We show that under MERA's RG transformations, the system will reach either the total reflection or the perfect transmission fixed point, depending on the sign of the interaction in the LL leads. Furthermore, we show that the correlation functions have a universal scaling behavior for attractive interactions. 
%For repulsive interactions, however, the scaling behavior depends on strength of the impurity. 
In addition, the boundary MERA provides crucial information about   the scaling dimensions for the primary fields in the CFT, which can be used to classify  RG fixed points.
%Thereby, with the ability to reach scale invariance under the MERA transformation in the presence of the impurity, and to perform a direct simulation of the semi-infinite chains, the boundary MERA, as we will demonstrate in the following, is  a promising scheme for extracting boundary effects of 1D quantum critical systems.

%These results, as we argue, is consistent with and reassure the conclusion of Ref.~\onlinecite{Kane1992}.  

The paper is organized as follows:
In Sec.~\ref{sec::introMERA}, we provide a brief review of the multi-scale entanglement renormalization ansatz.
In Sec.~\ref{sec:LL}, we discuss how to describe a two-wire junction as a Luttinger liquid with an impurity.
In Sec.~\ref{sec:bMERA}, we discuss how to construct the boundary Hamiltonian and how to obtain the boundary state from which   correlation functions and  scaling dimensions can be evaluated by optimizing a boundary MERA.
The current-current correlation functions at different RG fixed points are presented in Sec.~\ref{sec:JJ} and 
in Sec.~\ref{sec:SD} we show the spin-spin correlation functions and the scaling dimensions with and without the impurity.
Finally we summarize and discuss the advantage and the potential of the scheme in Sec.~\ref{sec:summary}.  Technical details on the implementation of the boundary MERA are presented in the Appendix. 

%%%%%%%%%%%%%%%%%%%%%%%%%%%%%%%%%%%%%%%%

\section{Multi-scale Entanglement Renormalization Ansatz}\label{sec::introMERA}
\begin{figure}
\centerline{
\includegraphics[width=8cm]{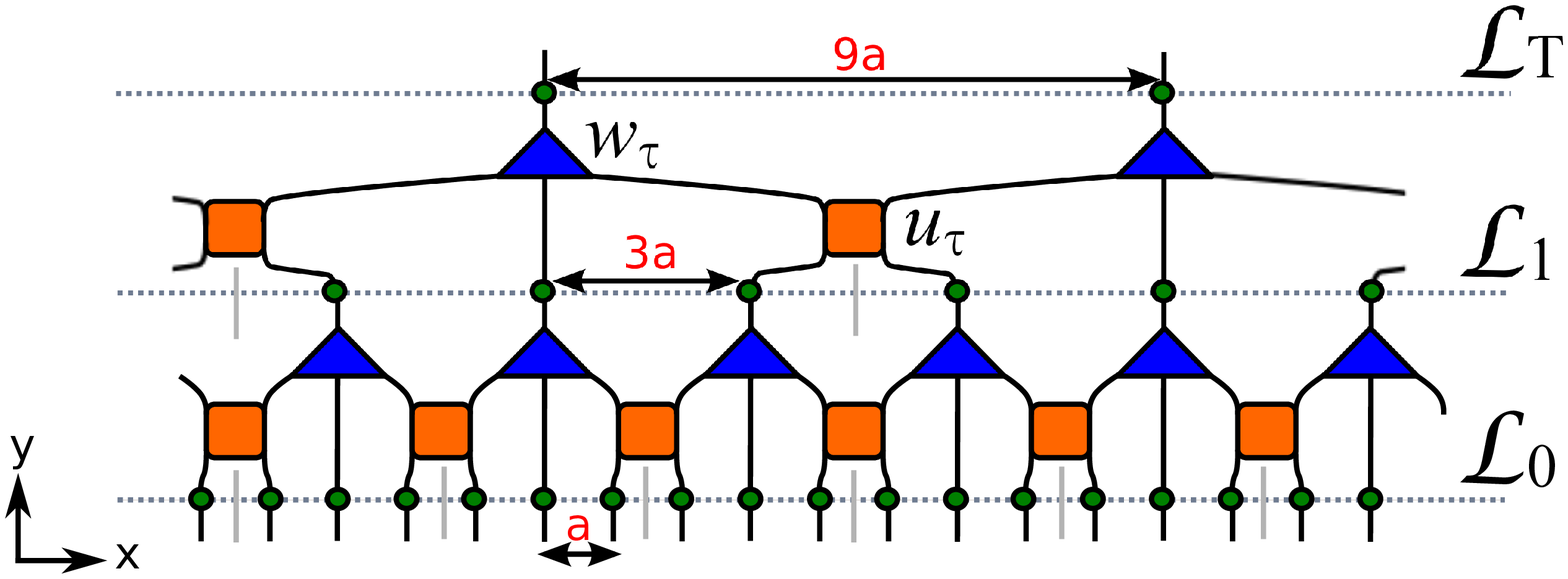}}
\caption{(Color online) Ternary MERA with periodic boundary condition for lattice length $N=18$ lied in x-axis. Blue triangles are isometries $w_\tau$, and orange squares are disentanglers $u_\tau$. $\mathcal{L}_\tau$ indicates various lattice layers with $\tau=0,1,T$. The gray vertical lines separate three lattice sites with green circles as one foundation block in each lattice layer. Each lattice spacing  is renormalized by the corresponding RG transformation, and the length scales of effective lattices are changed along the y-axis.}
\label{fig:3MERA_setup}
\end{figure}
In this section, we give a brief review of the basic concepts and properties of the MERA  tensor network, and we refer the readers to Ref.~\onlinecite{MERAalgorithm} and references therein  for more details.
  
\begin{figure}
\centerline{
\includegraphics[width=0.6\columnwidth]{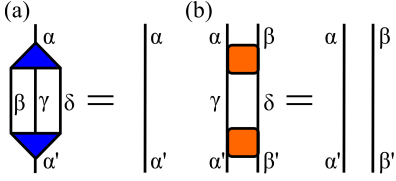}}
\caption{(Color online) Diagrammatic representation of the constraints for the isometries $w$ and disentanglers $u$ of the ternary MERA.}\label{fig:constraint}
\end{figure}

The MERA is a flexible real space RG scheme based on the tensor network and is designed to retain only the long-range entanglement of the system.~\cite{ER,MERACFT,MERABCFT,MERAalgorithm,Glen13_2,WengER,FazioER} This makes MERA an ideal method for simulating quantum critical systems with divergent correlation lengths.
%Starting from the original lattice Hamiltonian at layer $\mathcal{L}_0$ the MERA uses specific tensor network to transform the the Hamiltonian at layer $\mathcal{L}_0$ 
%into an effective Hamiltonian with coarse-grained lattice sites at layer $\mathcal{L}_1$.
%By repeating the same procedure new effective Hamiltonians are generated at more and more coarse-grained lattice sites at layers $\mathcal{L}_\tau$.
%If the original lattice at layer $\mathcal{L}_0$ is finite then after a finite number of RG transformations the coarse-grained lattice $\mathcal{L}_\tau$
%will become small enough to be treated exactly. However if the original lattice at layer $\mathcal{L}_0$ is infinite then one can perform the RG transformation indefinitely, 
%but after some layer $\mathcal{L}_\tau$ the system will become scale invariant.}
In this work we adopt the ternary MERA scheme where three lattice sites at $\mathcal{L}_\tau$ are coarse-grained into a single site at $\mathcal{L}_{\tau+1}$.
%The number of sites at $\mathcal{L}_{\tau+1}$ is hence one-third of the number of sites at $\mathcal{L}_\tau$.
In  \fig{fig:3MERA_setup} we illustrate the ternary MERA scheme with the $N=18$ sites subjected to a periodic boundary condition. 
The top lattice layer $\mathcal{L}_T$ with two sites are obtained via two RG transformations,
\begin{align}
\mathcal{L}_0 \stackrel{\rm RG}{\rightarrow} \mathcal{L}_1 \stackrel{\rm RG}{\rightarrow} \mathcal{L}_T .
\end{align}
The ternary MERA scheme is consist of two major ingredients: 
(1) The \textit{disentangler} $u_\tau$ that removes the short range entanglement within the corresponding length scales. 
(2) The \textit{isometry} $w_\tau$ that merges three sites at layer $\mathcal{L}_\tau$ to form one site at layer $\mathcal{L}_{\tau+1}$. 
During the simulation the $u_\tau$ and $w_\tau$ are optimized iteratively based on the variational principle.
It is essential that one first applies the disentangler $u_\tau$ before applies the isometry $w_\tau$ to merge lattice sites.
Another key feature of the MERA scheme is that  the isometry $w_\tau$ and the unitary  $u_\tau$ must satisfy the constraints, (Fig.~\ref{fig:constraint})
\begin{align}
&\sum_{\beta\gamma\delta} (w_\tau)^{\alpha}_{\beta\gamma\delta} (w_\tau^\dagger)^{\beta\gamma\delta}_{\alpha'}  = \delta_{\alpha\alpha'},
\\
&\sum_{\gamma\delta} (u_\tau)^{\alpha\beta}_{\gamma\delta} (u_\tau^\dagger)^{\gamma\delta}_{\alpha'\beta'} =\delta_{\alpha\alpha'}\delta_{\beta\beta'}.
\end{align}
These  constraints ensure that local operators are transformed into local operators and make possible to evaluate two-point correlation functions within the MERA framework.

For an infinite lattice $\mathcal{L}_0$ one can perform infinite many RG transformations, resulting in a MERA tensor network   similar to \fig{fig:3MERA_setup} with  infinite  sites and  layers. Assuming  translational invariance,  a single pair of $(u_\tau, w_\tau)$ is enough to uniquely define the coarse-graining process into $\mathcal{L}_\tau$.
For critical systems,  after a finite number of RG transformations, the system becomes scale invariant at $\mathcal{L}_s$.
After the system reaches  scale invariance, further RG transformation will generate the same effective Hamiltonian and the lattice.
Hence it suffices to use a single pair of $(u_s, w_s)$ to represent the RG transformations for $\mathcal{L}_\tau$ for $\tau \ge s$.
%In other words, one has $(u_\tau, w_\tau)=(u_s, w_s)$ for  $\tau \ge s$. 
Such a MERA structure is called scale invariant MERA in the literature.\cite{MERACFT}
In principle, the number of RG steps to reach scale invariance 
is a priori unknown and depends on the original Hamiltonian. In practice to keep the computation trackable one sets $s$ to some pre-determined number
and  layers $\mathcal{L}_\tau, \tau=1, \cdots, s-1$ are called buffer layers in MERA terminology.

An advantage of the scale invariant MERA is its ability to directly extract scaling properties of a critical system. For example, scaling dimensions of primary fields 
and the central charge of the corresponding CFT can be obtained directly.~\cite{MERACFT} At scale invariant layers the RG transformation of operators
is dictated by the scaling superoperator $\widetilde{S}$ which is a fixed-point RG map. The scaling operator $\phi_i$ with scaling dimension $\Delta_i$ should satisfy the equation
\begin{align}
\label{eq:scaling_fun}
\widetilde{S} (\phi_i) = \lambda_i \phi_i\;,\;\;\Delta_i \equiv -\log_3 \lambda_i\;,
\end{align}
where the logarithmic base three reflects the three-to-one coarse-graining of the ternary MERA scheme.  
All scaling dimensions can be, in principle, obtained by evaluating eigenvalues of the superoperator.

\begin{figure}
\centerline{
\includegraphics[width=8cm]{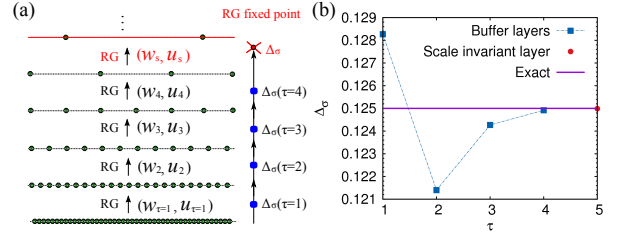}}
\caption{(Color online) (a) Four pairs of $(w_\tau,u_\tau)$ in buffer layers $(\tau=1,2,3,4)$  and the same copies of the scale invariant pair $(w_s,u_s)$ in the scale invariant layers are employed. Along the RG transformation axis, the pseudo-scaling-dimension $\Delta_\sigma(\tau)$ of buffer layers flows into the real scaling dimension $\Delta_\sigma$ of the scale invariant RG fixed point. (b) Both $\Delta_\sigma(\tau)$ and the scaling dimension of the spin primary field in the CFT are calculated, and the pseudo-scaling-dimension approaches the exact value as the layer $\tau$ is  close to the scale invariant layers. The exact value of the scaling dimension of the spin primary field  from the CFT is $1/8$ for the transverse Ising model.~\cite{Cardy2010} The bound dimension used here is $\chi=8$. }
\label{fig:s1buf5}
\end{figure}

As an example, we show the results of scaling dimensions for the 1D  transverse Ising model at criticality. We set $s=5$ as the scale invariant layer, making $\tau=1, \cdots, 4$ be the buffer layers.
We first optimize $(u_\tau, w_\tau)$ and $(u_s, w_s)$ by the standard MERA algorithm. The superoperator $\widetilde{S}$ is then constructed from  $(u_s, w_s)$ and diagonalized to obtain the (lowest few) scaling dimensions.
One can also construct superoperator $\widetilde{S}_\tau$ from $(u_\tau, w_\tau)$, although the system  is not yet scale invariant.  In the same token, pseudo-scaling-dimensions $\Delta_i(\tau)$ for the buffer layers can be obtained by 
diagonalizing $\widetilde{S}_\tau$. These $\Delta_i(\tau)$'s are used to monitor how the system approaches the scaling invariance. 
Thereby, we calculate $\Delta_i(\tau)$ for each buffer layer to form a flow of the pseudo-scaling-dimension as illustrated in \fig{fig:s1buf5}(a). 
We observe that the pseudo-scaling-dimension $\Delta_1(\tau)$ of buffer layers gradually approaches the value of the scaling dimension $\Delta_1$ of the scale invariant layers for $\tau \ge 5$.
The exact value of $\Delta_1=1/8$ is also plotted as a reference.
The above example shows that the scale invariant MERA provides a well-defined method to study the scaling properties of the RG fixed points. In the following, we will use this scheme to study the scaling properties of the junction of two interacting quantum wires. 
%The RG flow before reaching scale invariance can be monitored
%via the super-operators $\widetilde{S}_\tau$ in the buffer layers. The scaling dimensions of the scaling operators can be obtained by the scaling super-operator $\widetilde{S}$ in the scaling layers.
%The scaling super-operator $\widetilde{S}$ is constructed from $(u_s, w_s)$ which are both rank four tensors with four legs and have the same bond dimension $\chi$ for each leg.
%Therefore, the maximum number of the degree of freedom is $\chi$ for a single site operator and is $\chi^2$ for a two-site operator. 

%%%%%%%%%%%%%%%
%%%%%%%%%%%%%%%
%%%%%%%%%%%%%%%
%%%%%%%%%%%%%%%

%and we show a specific optimization concept in~\App{SEC:Optimization} for one boundary case.

\section{Junction of two interacting quantum wires} 
\label{sec:LL}

We start by modeling the impurity as a junction linking two identical semi-infinite 1D wires with a total Hamiltonian $H=H_w+h_B$. Here, $H_w$ represents the lattice Hamiltonian of the wires at half-filling, 
\begin{align}
  \label{eq:H-2-wires}
  H_{\rm w}=\sum_{\mu \in I, II} \sum_{i=0}^{\infty} 
  \left(- c^{\mu\dag}_{i+1} c^{\mu}_{i} + h.c. +V \bar{n}^{\mu}_{i} \bar{n}^{\mu}_{i+1} \right),
\end{align}
while the hopping Hamiltonian at the junction is given by
\begin{align}
  \label{eq:H-boundary}
  h_{\rm B}= - t \left( c^{I\dag}_{0} c^{II}_{0} + h.c. \right).
\end{align}
We denote $c^{\mu}_{i}$ ($c^{\mu\dag}_{i}$) with $\mu \in I, II$ as the annihilation (creation) operator at the site $i$ of the wire $\mu$, $\bar{n}^{\mu}_{i}\equiv c^{\mu\dagger}_{i} c^{\mu}_{i}-1/2$, and $V$ as the nearest-neighbor interaction strength. Following the bosonzination scheme,~\cite{Oshikawa2006} the wires can be represented in terms of continuum bosonic fields $\varphi^{\mu}$ and their dual fields $\theta^{\mu}$ by
\begin{equation}
  \label{eq:Hw-continuum}
  H_{\rm w}(\varphi^\mu,\theta^\mu) 
  = \sum_{\mu\in I,II} \frac{v}{4\pi}\int dx \left[g (\partial_x \varphi^\mu )^2+\frac{1}{g}(\partial_x \theta^\mu)^2\right],
\end{equation}
where, in the range $|V|\le 2$, the plasmon velocity $v$ and the Luttinger parameter $g$ are identified via the Bethe Ansatz at half filling as
\begin{equation}
  \label{eq:bethe-asatz-relation}
  v=\pi \frac{\sqrt{1-(V/2)^2} }{\arccos (V/2)},\quad g=\frac{\pi}{2 \arccos(-V/2) }.
\end{equation} 
Hence, we have $g=1$ for noninteracting wires and $g<1$ ($g>1$) for repulsive (attractive) interactions.

In comparison with an infinite LL wire, the presence of the junction could change the scaling behavior of the correlation functions across the junction. Starting from the lattice operators, define 
the current operator $J^{\mu}_{j+\frac{1}{2}}$ and the fermion density operators $N^{\mu}_{j+\frac{1}{2}}$ 
on the bond between sites $j$ and $j+1$ as 
\begin{align}
  \label{eq:total-current-density-lattice}
  J^{\mu}_{j+\frac{1}{2}} 
  = & i \left(c^{\mu\dag}_{j+1} c^{\mu}_{j} - c^{\mu\dag}_{j} c^{\mu}_{j+1}\right), \\
  N^{\mu}_{j+\frac{1}{2}} 
  = &  \frac{1}{2} \left(n^{\mu}_{j} + n^{\mu}_{j+1} - \langle n^{\mu}_{j} \rangle - \langle n^{\mu}_{j+1}\rangle \right). \nonumber
\end{align}
With these lattice operators, two-point correlation functions, such as $\langle J^{I}(x) J^{II}(x) \rangle$ and $\langle N^{I}(x) N^{II}(x) \rangle$, can be evaluated using the boundary MERA. The evaluated correlation functions should exhibit power law decay as expected in a 1D scale invariant quantum critical system. To see how the CIBC emerges due to the presence of the impurity at the RG fixed point, it is useful to introduce incoming and outgoing chiral density operators $\rho^{\mu}_{in/out}(x)$, defined with respect to the junction,  with the relations $J^{\mu}(x)= v (\rho^{\mu}_{out}(x)-\rho^{\mu}_{in}(x))$ and $N^{\mu}(x)= \rho^{\mu}_{out}(x)+\rho^{\mu}_{in}(x)$. It is worth to emphasize that these chiral densities are those diagonalizing the interacting Hamiltonian in Eq.~\eqref{eq:Hw-continuum} but not the chiral currents defined in the non-interacting bands.   

Since the boundary condition will dictate both the long distance scaling behaviors and the amplitude of correlation functions of primary fields,~\cite{Cardy1991} the chiral density correlation functions change accordingly with respect to different CIBC.~\cite{Rahmani2010} We can now decompose the two-points correlation functions with operators defined in Eq.~\eqref{eq:total-current-density-lattice} to obtain the chiral density correlation functions. For instance, we have, in the case of $\mu\neq \mu'$,
\begin{equation}
  \label{eq:chiral-cc-decomposed}
  \langle \rho^{\mu}_{out}(x) \rho^{\mu'}_{in}(x) \rangle  
  =- \frac{1}{2}\left(\frac{1}{v^2} \langle J^{\mu}(x) J^{\mu'}(x) \rangle 
  + \frac{1}{v} \langle N^{\mu}(x) J^{\mu'}(x) \rangle\right), 
\end{equation}
where we have used $\langle \rho^{\mu}_{out(in)} \rho^{\mu'\neq \mu}_{out(in)}\rangle=0$. In the presence of time reversal symmetry (which is our case), the second term in Eq.~(\ref{eq:chiral-cc-decomposed}) always vanishes. Thereby, the chiral correlation functions between different wires are directly proportional to the current-current correlation function.

Since the bulk of the LL quantum wires remains conformal invariant in the presence of impurity, correlation functions, in general, follow power law behaviors. Therefore, we expect that the equal time current-current correlation function decays at long distance in the form
\begin{align}
\label{eq:def-A-alpha}
  \left| \left\langle J^\mu(x) J^{\mu'} (x) \right\rangle \right| & \sim A x^{-\alpha}\;,
\end{align}
for $\mu \neq \mu'$. From RG prospect, the tunneling term between two LL wires is a relevant perturbation for attractive interactions, $g>1$, and is irrelevant for repulsive interaction, $g<1$. As a result, two semi-infinite LL wires effectively fuse into a single infinite LL wire at RG fixed point for $g>1$.~\cite{Kane1992} In this case, the leading contribution to the correlation function in Eq.~\eqref{eq:def-A-alpha} is universal regardless of the impurity strength, and has the prefactor $A= g v^2/8\pi^2$ and the exponent $\alpha=2$, c.f. Appendix~\ref{SEC:correlation}.~\cite{Rahmani2010,Rahmani2012} Thus, for $g>1$ the stable RG fixed point is a perfect transmission RG fixed point.

On the other hand, another RG fixed point corresponds to two disconnected wires with a strict zero linear conductance for $g<1$. An immediate consequence of this fixed point is the vanishing of $1/x^2$ term for the current-current correlation function in Eq.~\eqref{eq:def-A-alpha}. However, subleading contribution can come from the irrelevant boundary operators, which gives a faster power law decay with the exponent $\alpha>2$ and the prefactor depending on the strength of the impurity. Here, the exponent is non-universal and can be contingent on the detail of the impurity.

%\textbf{In the Kane-Fisher problem,  for attractive interactions $(g>1)$,  the impurity should heal at the RG fixed point and the two wires fuse into a single LL; therefore the current-current correlation function  decays  at long distance as a power law
%\begin{align}
% \left| \left\langle J^\mu(x) J^{\mu'} (x) \right\rangle \right| &= A x^{-\alpha}\;,
%\end{align}
%where $\mu \neq \mu'$. The prefactor $A= g v^2/8\pi^2$ and the exponent $\alpha=2$ are universal regardless of the impurity strength. On the other hand, for repulsive interactions $(g<1)$, the RG fixed point corresponds to two disconnected wires with zero conductance; therefore, the coefficient in  front of the leading term $1/x^2$ vanishes. There are subleading corrections coming from the irrelevant boundary operators, so  we expect to have a non-universal exponent $\alpha>2$ such that the current-current correlation function decays faster than $1/x^2$, and  the prefactor depends on the strength of the impurity. }

In the linear response regime, the chiral correlation functions in Eq.~\eqref{eq:chiral-cc-decomposed} can be used to determine the conductance across the impurity. From the conventional Kubo formula,~\cite{Oshikawa2006}
\begin{multline}
\label{eq:Kubo}
 G_{\mu\nu} = \lim_{{\omega} \rightarrow 0_+}-\frac{e^2}{\hbar}\frac{1}{
	{\omega} L}\int_{-\infty}^\infty d\tau\;e^{i {\omega}\tau}
\\
\times \int_0^L dx \;\langle {\cal T}_\tau J^\mu(y,\tau) J^\nu(x,0)\rangle,
\end{multline}
the imaginary-time ordered (indicated by ${\cal T}_\tau$) dynamical current-current correlation function for currents $J^\mu$ and $J^\nu$ on wires $\mu$ and $\nu$ is needed to evaluate the conductance. As the current operators can be represented in terms of the chiral density operators, we can decompose the non-chiral correlation function by chiral current correlation functions. For $\mu\neq \nu$, we have 
\begin{multline}
\label{eq:cc-function-chiral-decomposed}
\langle {\cal T}_\tau J^\mu(y,\tau) J^\nu(x,0)\rangle =
\\
 - v^2 \left( \langle \rho^{\mu}_{out}(y,\tau) \rho^{\nu}_{in}(x,0)\rangle + \langle \rho^{\mu}_{in}(y,\tau) \rho^{\nu}_{out}(x,0) \rangle \right),
\end{multline}
where we have used the fact that correlation functions vanish for the same chiral current in different wires. In the presence of the conformal symmetry and the CIBC, one can show that the chiral correlation functions in Eq.~\eqref{eq:cc-function-chiral-decomposed} is always a function of $z= v \tau \mp \mathrm{i} (x+y)$.~\cite{Rahmani2010} As a result, the dynamical chiral current correlation functions can be reconstructed via the static correlation functions shown in Eq.~\eqref{eq:chiral-cc-decomposed}. Finally, the fixed-point conductance can be subsequently evaluated using the Kubo formula in Eq.~\eqref{eq:Kubo}.

%%%%%%%%%%%%%%%%%%%%%%%%%%%%%%%%%%%%%%%%
\section{Boundary MERA} 
\label{sec:bMERA}

%For two-wires without a junction, the current-current correlation function at fixed points can be estimated by the bulk scale invariant MERA associated with the wires. On the other hand, an impurity can dramatically change the behavior of systems and leads to different fixed points. 

%$\bullet$ {\bf A paragraph that explain the boundary MREA and CFT boundary state}
%A boundary state reveals the scale invariant behavior at the fixed point from  boundary CFT point of view~\cite{Cardy1991}.

The boundary CFT predicts that each boundary RG fixed point is associated with a CIBC and hence a conformally invariant boundary state.~\cite{Cardy2010} As a result,  scaling behavior of the boundary operators are directly controlled by the realized boundary condition. In addition, even though the scaling dimensions of bulk primary operators, such as the chiral current operators, remain unchanged in the presence of a boundary, the coefficients of their correlation functions are dictated by the given boundary state. Thereby, constructing the corresponding boundary state allows us to obtain the full properties of a junction at its RG fixed point.~\cite{Cardy1991} In this section, we will discuss how to obtain the boundary state using a numerical boundary MERA scheme.  For a complete review of the MERA algorithm and detailed discussion on the MERA with impurities, we refer the interested readers to Refs.~\onlinecite{ER,Glen13_1,Glen13_2}.

%%%%%%%%%%%%%%%%%%%%
\begin{figure}[tbp]
  \centerline{\includegraphics[width=8cm]{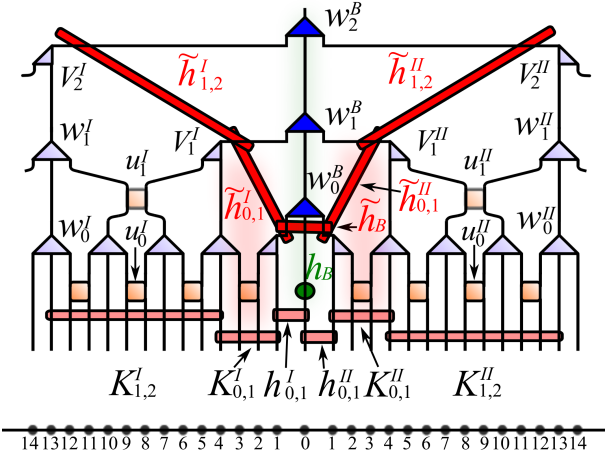}}
  \caption{\label{fig:defectLatt} 
  (Color online)
  Sketch of the boundary scale invariant MERA structure for three layers.
  The central tensors $w_\tau^B$ are used to represent the boundary state inside the causal cone (light green shaded area) of the green impurity.  
  $\widetilde{h}^\mu_{\tau,\tau+1}$ is an effective two-site boundary Hamiltonian 
  which is obtained by inhomogeneous coarse-graining,  
  related to the causal cone of $K^\mu_{\tau,\tau+1}$ such as the red shaded area.
  Here $w^\mu_\tau$, $u^\mu_\tau$, and  $V^\mu_\tau$ are bulk isometries , bulk disentanglers, and boundary truncation tensors respectively.
  The effective impurity Hamiltonian $\widetilde{h}_B$ is constructed from  the impurity  Hamiltonian $h_B$ and  two-site Hamiltonian $h^I_{0,1}$ and $h^{II}_{0,1}$ in the bulk wires.
  Two quantum wires are connected in the junction, and we also label the site indices for each wire.
%  Here $\mu\in I,II$ corresponds to the left and right wires, respectively. 
  } 
  \label{fig:defectMERA}
\end{figure}
%%%%%%%%%%%%%%%%%%%%

In \fig{fig:defectMERA} we sketch the MERA structure that describes two semi-infinite wires with a junction. First, two sets of standard bulk scale invariant MERA with isometries $w^\mu_\tau$ and disentanglers $u^\mu_\tau$ of bond dimension $\chi$ are used to describe the two semi-infinite wires. Second,  the bare Hamiltonian at the original lattice are regrouped into $K^\mu_{\tau,\tau+1}$ and  inhomogeneously ascended using the truncation tensors $V^\mu_\tau$ with bond dimension $\chi^B$, $w^\mu_\tau$, and $u^\mu_\tau$  to form the boundary Hamiltonian $\widetilde{H}_B$ . Finally the  central tensor $w^B_\tau$ is used to describe the boundary state and is optimized via the boundary Hamiltonian $\widetilde{H}_B$. In a nutshell, the scale invariant boundary state is represented by the scale invariant central tensor  $w^B_\tau$ in the boundary MERA.  In the following we summarize the  major steps of the boundary MERA algorithm, and we refer the readers to the Appendix for more details:

%In the nutshell, the scale invariant boundary state is represented by central tensors as shown in \fig{fig:defectMERA} for the boundary MERA. The boundary MERA provides a well defined procedure that finds the matrix-product-state (MPS) representation of the ground state for an effective coarse-grained boundary Hamiltonian $\widetilde{H}_{B}$ that entangles the boundary and bulk Hamiltonian. In what follows, we detail the boundary MERA approach in three major steps: 
%\begin{enumerate}

%\item 
\textit{Optimization of the bulk scale invariant MERA} --
MERA is a specific scheme to perform real-space RG transformations using isometries $w^\mu_\tau$ (light blue triangles) and disentanglers $u^\mu_\tau$ (yellow squares) as shown in \fig{fig:defectMERA}.~\cite{MERAalgorithm} In each RG step, to construct the coarse-grained Hamiltonian at the next layer $\tau+1$,  the disentangler $u^\mu_\tau$ is used to transform to a less entangled local basis between blocks while the isometry $w^\mu_\tau$ is used to perform coarse-graining. They are optimized using  the bulk  scale invariant MERA algorithm.~\cite{MERAbook} The algorithm minimizes the energy per site  associated with the bare Hamiltonian,  shown as the light pink bars at the bottom of \fig{fig:defectMERA}.  In this step, each wire is treated as  independent   and the associated $u^\mu_\tau$ and $w^\mu_\tau$ are optimized independently.  In this work,  the two wires are identical, so the bulk optimization needs to be carried out only once.

\begin{figure}
\centerline{
\includegraphics[width=8cm]{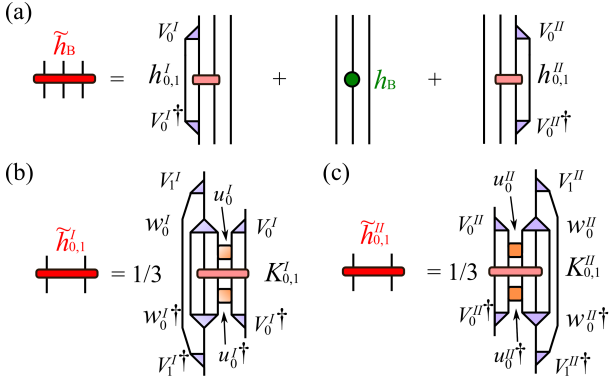}}
\caption{(Color online) (a) Graphic representation of $\widetilde{h}_B$ including the impurity Hamiltonian $h_B$ and the inhomogeneous coarse-graining of the first two-site Hamiltonian $h_{0,1}^\mu$ in wire $\mu\in I,II$.
(b)-(c) Using inhomogeneous coarse-graining to obtain the boundary Hamiltonian $\widetilde{h}_{0,1}^\mu$ with a scaling factor $1/3$ for  wire $I$ and $II$, respectively.}
\label{fig:ascend_k}
\end{figure}

\textit{Construction of the effective boundary Hamiltonian} --
A key step of the boundary MERA is to perform an  inhomogeneous coarse-graining of the bare Hamiltonian to obtain an effective boundary Hamiltonian $\widetilde{H}_B$.~\cite{MERABCFT} 
The boundary Hamiltonian for the chain of the central tensors $w^B_\tau$ consists of  the effective impurity Hamiltonian $\widetilde{h}_B$, pictorially defined in \fig{fig:ascend_k}~(a) and two-site  Hamiltonians $\widetilde{h}^\mu_{\tau,\tau+1}$ that connect two adjacent sites $\tau$ and $\tau+1$ as depicted as red bars in \fig{fig:defectMERA}:
\begin{align}
  \label{EQ:BDH}
  \widetilde{H}_B & = \widetilde{h}_B 
  + \sum_{\mu\in I,II} \sum_{\tau=0}^{\infty} \widetilde{h}^\mu_{\tau,\tau+1}.
\end{align}
Here, $\widetilde{h}^\mu_{\tau,\tau+1}$ is constructed from the inhomogeneous ascending of a collection $K^\mu_{ \tau,\tau+1}$ of bare Hamiltonians $h_{i,i+1}^\mu$ at the same scale
\begin{align}
  K^\mu_{ \tau,\tau+1} &=\sum_{i=s(\tau)}^{s(\tau+1)-1} h_{i,i+1}^\mu\;,
\label{EQ:K}
\end{align}
where  $s(\tau)=(3^{\tau+1}-1)/2$ with $\tau=0,1,2,3,\cdots$, and $ h_{i,i+1}^\mu$ is a two-site Hamiltonian in wire $\mu$. 
The boundary Hamiltonian $\widetilde{h}^\mu_{\tau,\tau+1}$ is obtained by  inhomogeneous coarse-graining two-site bulk Hamiltonian in layer $\tau$ via the inhomogeneous ascending superoperator $\overline{A}_{\rm bd}^\mu[h_{1,2}^\mu(\tau),h_{2,3}^\mu(\tau),h_{3,4}^\mu(\tau)]$, with a scaling factor $1/3$ which reflects the ternary MERA.  For instance, to obtain the boundary Hamiltonian $\widetilde{h}^\mu_{0,1}$, we construct $\overline{A}_{\rm bd}^\mu[h_{1,2}^\mu(\tau=0),h_{2,3}^\mu(\tau=0),h_{3,4}^\mu(\tau=0)]$ for $K_{0,1}^\mu$ in \fig{fig:ascend_k}~(b)-(c) by contracting the tensors inside the causal cone (shaded red area in \fig{fig:defectMERA}).   Moreover, in order to assign a different bond dimension $\chi^B$ to the central tensor $w^B_\tau$, we introduce a truncation tensor $V^\mu_\tau$  at  the boundary. 
%We refer the readers to the appendix  for details of the boundary truncation and the construction of the truncation tensor.
We note that
in general $\chi^B$ can be layer dependent until the central tensor $w^B_\tau$ reaches scale invariant, after which only one single $\chi^B$ is used for all the scale invariant layers. 
%We note that in order to represent the scale invariant fixed point, the boundary scale invariant MERA needs to have infinite layers. 
%We refer the readers to the appendix %\App{SEC:GetBdH} for more details.   
% the  procedures to construct the effective two-site boundary Hamiltonian $\widetilde{h}^\mu_{\tau,\tau+1}$, and . 

%%%%%%%%%%%%%%%%%%%%%%%%%%%%
  \textit{Optimization of the central tensors} -- 
 The final step is to utilize the boundary Hamiltonian (red bars) to optimize the central tensors $w_{\tau}^B$ (blue triangles) which represent the boundary state in the boundary chain (shaded green). Here we employ a scale invariant boundary MERA algorithm to optimize the central tensors $w^B_\tau$. Similar to the bulk MERA, we treat the energy per site  as the cost function for the optimization processes. The energy of the boundary MERA can be calculated at  layer $\tau$ as
\begin{align}
  E &= \tr \{ w_\tau^B Y_\tau \}\;,
  \label{EQ:Y}
\end{align} 
where $Y_\tau$ is the environment associated with $w_\tau^B$. In general, it is necessary to insert several buffer layers with different central tensors  $w_{0}^B, w_{1}^B, \cdots, w_{\tau_{{\it bf}-1}}^B$ before one reaches the scale invariant layers  characterized by a single central tensor  $w_{s}^B$. For the buffer layers and the scale invariant layers the environment construction differs. The environment for the former can be obtained by the procedure defined in \App{SEC:Optimization}. The environment for the scale invariant layers is constructed from the scale invariant Hamiltonian
\begin{align}
  \widetilde{h}_s^\mu = \sum_{\tau=\tau'_s}^\infty \frac{\widetilde{h}_{\tau,\tau+1}^\mu}{3^{\tau-\tau'_s}}\;,
\label{EQ:HS1}
\end{align}
where  the layer $\tau$ starts from the second scale invariant layer $\tau'_s=\tau_s+1$, and all layers beyond  layer $\tau_s$ are scale invariant. Here the factor three reflects the three-to-one coarse-graining. In practice, it is useful to introduce a cutoff to replace the infinite sum by a finite sum (see Appendix). 
Given the  effective Hamiltonian $\widetilde{H}_B$, 
%one can also try to represent the boundary state by an MPS~\cite{MERABCFT}  and use variational methods or DMRG to optimize the MPS. 
%However, it might be difficult to reach RG fixed points represented by a scale invariant MPS.
%However, it would be hard to reach scale invariance if MPS based methods are used. 
%Here 
we perform an optimization procedure based on the boundary MERA framework.~\cite{MERABCFT} 
The procedure is similar to optimizing the scale invariant  MERA and allows us to construct an scale invariant boundary state.
We describe the details of the construction of the boundary MERA tailored for the two-wire junction in the  Appendix. 
%A  more general scheme is  available in Refs.~\onlinecite{Glen13_1,Glen13_2}. 
%For  clarity, we introduce the construction of the tensor network  for  the boundary MERA  following  Ref.~\onlinecite{Glen13_2}.
In all calculations below we always set to have two buffer layers and enforce the system to have scale invariance starting from the third layer.

%reach the scale invariance.

%Here, we design an optimization procedure for the boundary MERA, similar to that for the bulk MERA, to reach the scale invariance. In the MERA framework, the one-dimensional system is mapping onto (1+1) dimension problem, and the extra dimension indicates the various length scales. The scale invariant symmetry is executed in the extra dimension by using the copies of $w_s^B$ as the central tensors of scale invariant layers. Comparing with other numerical methods, MERA characterizes the properties at fixed points very well, because the scale invariant symmetry is built-in for the MERA. 
%In the boundary MERA, the scale invarinat central tensors are the same due to the scale invariant character, and the set of central tensors are handled as variational parameters, optimized by minimizing the cost function of the boundary Hamiltonian.

%%%%%%%%%%%%%%%%%%%%
%\begin{figure}[htbp]
 %\centerline{\includegraphics[width=9cm]{g15_g124_JJ_out}}
 % \caption{The current-current correlation function as a function of the distance from the boundary. It shows universal properties  with various $t=0.5,0.8,0.9$ for  $g>1$.
% (a) $A\sim 0.032,\alpha \sim -2$ for $g=1.5$. 
 %(b) $A\sim 0.04,\alpha \sim -2$ for $g\sim1.24$. }
 % \label{fig:bMERA}
%\end{figure}

\section{Current-current correlation functions}
\label{sec:JJ}
%One can also carry out the MERA calculation in the fermion picture.
As stated previously, the current-current correlation functions across the junction provide important information on the transport properties. 
To simplify the calculations,  we perform a Jordan-Wigner transformation to 
map the spinless fermion model into a spin-1/2 XXZ model. 
We consider two semi-infinite wires, labeled by $\mu=I,II$, and the transformation is defined as 
\begin{align}	
c^\mu_j &= S^{\mu -}_j e^{\ii \pi \Phi_j^\mu}\;,\\
c^{\mu \dagger}_j &= S^{\mu +}_j e^{-\ii \pi \Phi_j^\mu}\;.
\end{align}
The site index $j$ goes from zero to infinity in each wire, and
the junction connecting the two wires is at site zero (see \fig{fig:defectMERA}).
In addition, the phase factor $\Phi_j^\mu$ is defined as 
\begin{align}
\Phi_j^{I} &= \sum_{k=\infty}^{j+1}S^{I +}_k S^{I -}_k\;,\\
\Phi_j^{II} &= \sum_{k=\infty}^{0}S^{I +}_k S^{I -}_k 
+  \sum_{k=0}^{j-1}S^{II +}_k S^{II -}_k \;.
\end{align}
%where for the wire-$I$, $k$ runs from infinity to $(j+1)$ while for the wire-$II$, $k$ is small than $j$.
Additionally the current operator in \eq{eq:total-current-density-lattice} in the spin language is written as
\begin{align}
 J_{j+\frac{1}{2}}^\mu = \ii (S^{\mu -}_{j} S^{\mu +}_{j+1} - S^{\mu +}_{j}S^{\mu -}_{j+1})\;.
\end{align}

Once an optimal boundary MERA state is obtained, we can evaluate the current-current correlation function
in the presence of the junction.
For the lattice model, we calculate $\left| \left\langle J^{\mu}_{j+\frac{1}{2}} J^{\mu'}_{j+\frac{1}{2}} \right\rangle\right|$, 
where $j+\frac{1}{2}$ denotes the current from  the $j$-th to the $(j+1)$-th site. $x=(j+\frac{1}{2})$ defines 
the distance from the boundary.  
%For the Kane-Fisher problem, one expects that the current-current correlation function decays 
%at long distance as a power law
%\begin{align}
%  \left\langle J^\mu(x) J^{\mu'} (x) \right\rangle &= A x^{\alpha}\;,
%\end{align}
%where $\mu \neq \mu'$, and the measurement of $\langle J^\mu(x) \rangle$ should be zero at the boundary.

%%%%%%%%%%%%%%%%%%%%
\begin{figure}[htbp]
 \centerline{\includegraphics[width=9cm]{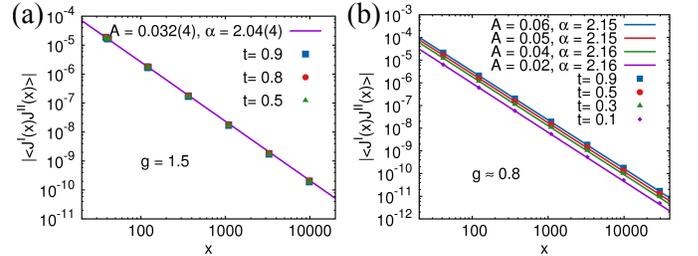}}
 \caption{(Color online) The current-current correlation function as a function of the distance from the boundary. 
 (a) Universal behavior with the same $A=0.032,\alpha \approx 2$ for $g=1.5$ with various $t=0.5,0.8,0.9$ is observed. 
 (b) Non-universal behavior with distinct $A$ and $\alpha > 2$ for $g \approx 0.8$ and $t= 0.9, 0.5, 0.3, 0.1$.
 The calculations are carried out using $\chi=12$ and $\chi^B=24$.
  }
\label{fig:bMERA}
\end{figure}

%%%%%%%%%%%%%%%%%%%%

% For g>1, we have $A= -g v^2/8\pi^2$ and $\alpha =2$. Here, $g$ and $v$ are given in Eq. (4). For g<1, however, A becomes non-universal while $\alpha$ should depend on $g$. We do not have a good argument about what is the value of $\alpha(g)$. 

In the following we show our numerical results of current-current correlation function for $g=1.5$ and $g \approx 0.8$ as representatives for the $g>1$ and $g<1$ fixed points respectively.
For $g>1$ the RG fixed point corresponds to a healed single wire.
Furthermore, the boundary CFT predicts that the prefactor $A$  and the exponent $\alpha$ in Eq.~\ref{eq:def-A-alpha} are  universal regardless the strength of the junction.
In \fig{fig:bMERA}(a) we show the current-current correlation function for the case of $g=1.5$ and $t=0.5,0.8,0.9$ 
%over the distance $x \ge 13$. 
for large distance. 
We observe that  all data points fall on a universal line with the same exponent $\alpha=2.04(4)$ and the same prefactor $A=0.032$. 
These results agree well with the boundary CFT's prediction of $A= g v^2/8\pi^2$ using the velocity $v \approx 1.299$ from Eq.~(\ref{eq:bethe-asatz-relation}). 
For very short distance, we find that the current-current correlation functions depend on the coupling strength of the weak link.
We employ two buffer layers before we enforce scale invariance in our current MERA scheme; therefore,  for distance longer than a characteristic length scale $x_s=\frac{3^3-1}{2}=13$,  the correlation function is dictated by the RG fixed point and shows a universal behavior. For distance $x<x_s$, however, the the correlation function depends on the strength of the weak link.

In contrast, for $g<1$, the RG fixed point corresponds to two disconnected wires and the universal behavior is not expected.
Consequently, to the leading order, the coefficient in front of the correlation functions is zero and the sub-leading corrections from the irrelevant operators at the boundary will be observed.
Since the scale invariance of the MERA scheme is enforced, the correlation function will still show a power-law decay but with an exponent that is larger than 2 with a non-universal prefactor.
In  \fig{fig:bMERA}(b) we show the results for the case of $g \approx 0.8 <1 $ and $t=0.1,0.3,0.5,0.9$.
Indeed we observe that different $t$ results in different scaling behavior with the  exponents $\alpha> 2$. 
Similarly for short distance $x<13$ we also observe non-universal behavior since the system is not yet dictated by the RG fixed point.
The distinct behavior of the correlation function for $g>1$ and $g<1$  indicates that the system flows into different RG fixed points.
Even without the a priori knowledge about the analytical results for the number and the nature of the RG fixed points, the numerical results can distinguish the two RG fixed points.

Furthermore, the conductance for the two-wire model can be estimated by the Kubo formula using the current-current correlation function.\cite{Kane1992b,Rahmani2010} 
For $g>1$, we expect that the system is dictated by a total transmission fixed point, i.e., two wires are fused into a single LL wire. 
The exponent in the current-current correlation function is hence $\alpha=2$, leading to  the conductance 
\[
G_{I,II}=g\frac{e^2}{h}.
\]
On the other hand, for $g<1$, we expect the two wires are effectively disconnected, which corresponds to a total reflection fixed point. 
In this case,  the current-current correlation function between two wires should decay faster than $1/x^{2}$, resulting in a zero conductance.
Our results discussed above hence shows that one can use boundary MERA to classify fixed points from the exponent of the current-current correlation function. In the following section, we will show a more direct way to identify the fixed points using the scaling dimensions of the boundary operators.

%%%%%%%%%%%%%%%%%%%%%%%%%%%%%%%%%%%%%%%%%%%%%%%%%%%%%%%%%%%%
\section{Spin-spin correlation functions and scaling dimensions}
\label{sec:SD}
%%%%%%%%%%%%%%%%%%
%%%%%%%%%%%%%%%%%%
\begin{figure}
\centerline{\includegraphics[width=9cm]{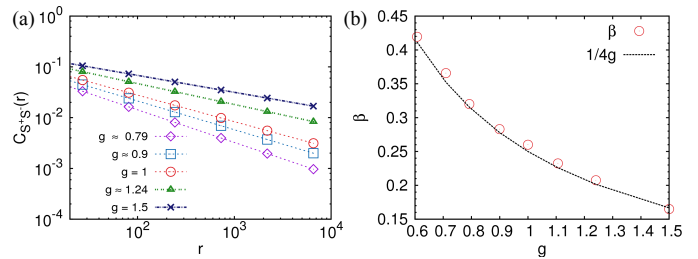}}
\caption{(Color online) (a) The spin-spin correlation function in~\eq{eq:tpcspsn} as a function of the distance between two spin operators for a bulk wire with $g \approx 0.79, 0.9, 1, 1.24, 1.5$. (b) The exponent $\beta$ of the spin-spin correlation function as a function of $g$. The calculations are carried out using $\chi=16$.}
\label{fig:bulk_spsn}
\end{figure}
%%%%%%%%%%%%%%%%%%
%%%%%%%%%%%%%%%%%%
We next study the two-point spin-spin correlation function defined as
\begin{align}
\label{eq:tpcspsn}
C_{S^+S^-} (r) &=\left| \langle S^+(r_1) S^- (r_2) \rangle - \langle S^+ \rangle \langle S^- \rangle  \right|\;, 
\end{align}
where $r_1$=$r_2$ is the distance from the spin operator to the impurity site on the wire $I$ and $II$ respectively. 
$r=|r_1|+|r_2|$ is the distance between two spin operators of $S^+(r_1)$ and $S^-(r_2)$. The correlation function shows a power-law decay,
\begin{align}
C_{S^+S^-} (r) =\alpha r^{-2\beta}\;,
\end{align}
where $\beta$ should be equal to the second lowest scaling dimension $\Delta_2$ of the bulk LL wire. 
The lowest scaling dimension $\Delta_1= 0$ which corresponds to the scaling operator of the identity, and is independent of the Luttinger parameter $g$. 
The lowest non-vanishing scaling dimension $\Delta_2$, however, varies with  $g$ as $1/4g$. In the spin language of the $XXZ$ model, this is the scaling dimension of the primary fields $S^\pm$,
leading to a power-law decay of the spin-spin correlation function. In~\fig{fig:bulk_spsn}(a), we plot the spin-spin correlation functions for several $g$ in the bulk wire.
We clearly observe that for all $g$'s, the spin-spin correlation functions show a power-law decay. In~\fig{fig:bulk_spsn}(b) we show the fitted exponent $\beta$ as a function of $g$.
The results agree well with the expected value of $\beta=\Delta_2=1/4g$. 
%Our results clearly demonstrate that MERA is an excellent tool extract the exponent of the power-law decay correlation functions. 
Since the exponents are dictated by the scaling dimensions of the primary fields, this provides an indirect way to study the scaling dimensions. 
We will demonstrate how to study  scaling dimensions directly in the MERA  later in this subsection.

\begin{figure}
\centerline{\includegraphics[width=9cm]{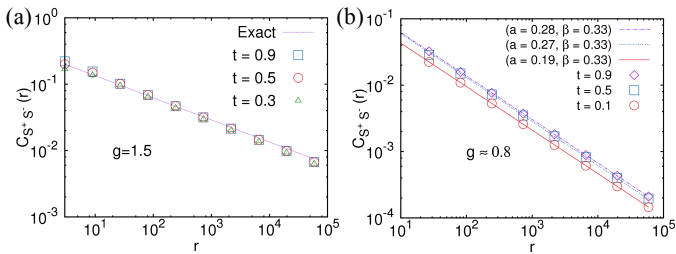}}
\caption{(Color online)  The spin-spin correlation function in~\eq{eq:tpcspsn} as a function of the distance between two spin operators crossing the junction  for various $g$, and  we chose two spin operators have same distance far away from the boundary. (a) The universal behavior at the  perfect transmission fixed point for $g=1.5$ with $t=0.3,0.5,0.9$,  the exact solution~\cite{Lukyanov1998,Lukyanov2003323} describes the character of the bulk. (b) The non-universal behavior at the total reflection fixed point for $g \approx 0.8$ with $t=0.1,0.5,0.9$, and the lines are the fitting results of $C_{S^+ S^-}(r) \approx a r^{-2\beta}$. }
\label{fig:SPSNG1508}
\end{figure}

To study the effects of the boundary we investigate how the behavior of spin-spin correlation function depends on the strength $t$ of the impurity.
In~\fig{fig:SPSNG1508}(a) and (b) we show the results for $g=1.5$ and $g\approx 0.8$ respectively with $t=0.1, 0.5$, and $0.9$.
For $g>1$, we again observe a universal behavior that all correlation functions fall on the same line regardless the strength $t$ of the impurity.
Furthermore, the line is actually the same as the one for the bulk wire. In contrast for $g<1$, non-universal behavior is observed.
The pre-factor $a$ depends strongly on the value of $t$. The exponent $\beta$, however, remains the same as the bulk value.
We comment that even without the a priori knowledge on the exact nature of the fixed points, the results obtained by MERA clearly indicate that there are two distinct fixed points
corresponding to the case of $g>1$ and $g<1$ respectively. For problems with unknown RG fixed points, in principle it is possible to identify the RG fixed points by
studying the behavior of different two-point correlation functions.

Another way to directly identify unknown RG fixed points is to study the scaling dimensions of the boundary scaling operators which can be straightforwardly obtained by boundary MERA.
Identifying operator contents of primary fields and their descendants are the most essential step to quantify the properties of a conformally invariant system. These \emph{scaling operators} $\phi_{\alpha}^{B}$ follow specific rule under the scaling transformation and have the scaling dimensions $\Delta_{\alpha}^B$.
Similar to the scale invariant bulk MERA~\cite{MERACFT}, in the scale invariant layer  the boundary scaling superoperator $\widetilde{\mathcal{S}}^{B}$ which can be expressed in terms of central tensor $w^{B}_s$ as

\begin{align}
\nonumber
  \left[ \widetilde{\mathcal{S}}^{B} \right]^{\alpha,\delta'}_{\delta,\alpha'} =& \sum_{\beta,\gamma} \left[ w^{B}_s\right]^\alpha_{\beta,\delta,\gamma} 
  \left[ {w_s^{B}}^\dag\right]_{\alpha'}^{\beta,\delta',\gamma}\;.
%\\=& \sum_{\beta,\gamma} \left[ w^{(B)}_s\right]^\alpha_{\beta,\delta,\gamma} 
% \left( \left[ {w_s^{(B)}}\right]^{\alpha'}_{\beta,\delta',\gamma}\right)^*.
\end{align}
Here, the upper index $^{B}$ indicate that superoperator is evaluated at the boundary. Then, one can show that the boundary scaling operators $\phi_\alpha^{B}$ are the eigen-operators of superoperator $\widetilde{\mathcal{S}}^{B}$ and have the relations~\cite{MERACFT,MERAalgorithm,MERABCFT}
\begin{equation}
  \widetilde{\mathcal{S}}^{B}(\phi_{\alpha}^{B}) = \lambda_{\alpha}^{B} \phi_{\alpha}^{B}, 
  ~~~~~\Delta_{\alpha}^{B} \equiv -\log_3 \lambda_{\alpha}^{B}.
  \label{eq:bulkS}
\end{equation}
The base three of the logarithm reflects the mapping of three sites into one during the coarse-graining. Now, the scaling dimensions $\Delta_\alpha^{B}$ of scaling operators are obtained simply by the eigenvalue decomposition of the $\widetilde{\mathcal{S}}^{B}$.  Numerically, with the finite boundary bond dimension $\chi^{B}$, the maximum number of scaling dimensions, which we can evaluate from boundary MERA, is constrained to be $(\chi^{B})^2$.

%%%%%%%%%%%%%%%%%%
%%%%%%%%%%%%%%%%%%
%%%%%%%%%%%%%%%%%%
%%%%%%%%%%%%%%%%%%

The boundary scaling dimensions are expected to show different dependence on the Luttinger parameter $g$, but are independent from the hopping amplitude $t$ at the the junction. First, we expect that $\Delta^{B}_2(g>1)= 1/4g$ has the same Luttinger liquid parameter dependence as that in the bulk. This is due to the fact that the boundary RG fixed point at $g>1$ corresponds to the perfect transmission between two wires and two semi-infinite LL wires effectively heal to one infinite LL wire. On the other hand, the RG fixed point for $g<1$ corresponds to a total reflection boundary condition for both wires. Due to the current conservation, the incoming current is perfectly reflected to the outgoing current at the  boundary for both wires. Therefore the current operators are pinned at boundary, i.e., $\theta^{I,II}|_{x=0}=0 \to \phi_{in}|_{x=0}=\phi_{out}|_{x=0}$, which lead to the change of scaling dimensions of boundary operators. (c.f. Ref.~\onlinecite{Oshikawa2006} for detailed arguments.) The spin operators at boundary now become $S^{\pm} \sim e^{\pm i \sqrt{g/2} \varphi} \to S^{\pm}|_{x=0}  e^{\pm \sqrt{2 g} \phi_{in}}$,~\cite{Oshikawa2006,Lukyanov1998} and have scaling dimension $\Delta^{B}_2= 1/ 2 g$.

\begin{figure}[htbp]
  \centerline{\includegraphics[width=8cm]{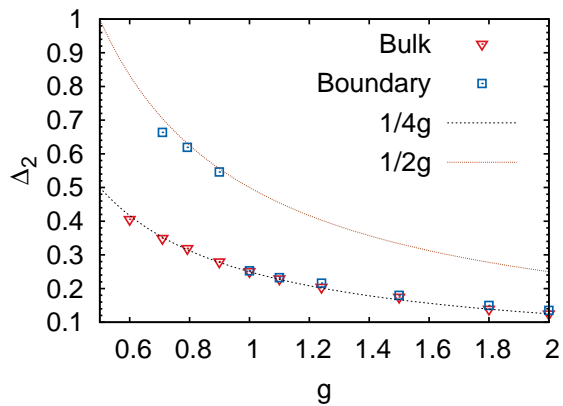}}
  \caption{ (Color online)
  The lowest few non-vanishing scaling dimensions of primary fields of the bulk and boundary as a function of Luttinger parameters $g$.
  The calculations are carried out using $\chi=12$ and $\chi^B=24$.
  }
  \label{FIG:fig3}
\end{figure}

The lowest non-vanishing bulk and boundary scaling dimensions are evaluated and shown as red triangles and blue squares in \fig{FIG:fig3}, respectively. First and foremost, the bulk scaling dimension fits very well with the expected functional dependence $\Delta_2= 1/4g$ while the boundary scaling dimension $\Delta_2^{B}(g)$ exhibits a drastic change at $g=1$. Numerically, we found $\Delta^{B}_2(g>1)= 1/4g$, the same Luttinger liquid parameter dependence as in the bulk. For $g<1$, we observed that the functional dependence of $\Delta^{B}_2(g<1)$ fits very well with $1/ 2 g$. These results are consistent with fact that two different boundary RG fixed points are realized at $g>1$ and $g<1$.

%{\bf Do we need to elaborate more about the numerical parameters here? say $\chi$ and some other details.} 

%%%%%%%%%%%%%%%%%%%%
%%%%%%%%%%%%%%%%%%%%

%%%%%%%%%%%%%%%%%%%%%%%%%%%%%%%%%%%%%%%%%%%%%%%%%%%%%%%%%%%%
\section{Conclusions}
\label{sec:summary}

%In conclusion, we use boundary MERA to classify the fixed points of the impurity system, and we get two fixed points for this impurity problem. One fixed point shows universal properties in that we get the same coefficient $A$ for $g>1$.  The other shows non-universal properties and systems' details would show in the coefficient $A$ for $g<1$.  Boundary MERA maybe an efficient method to classify fixed points and to study the behavior of RG fixed points. Moreover, this method can be extend to a junction of M quantum wires, and  fixed points of M-wires can be classified.  In addition, we numerically compute the four-points correlation function related to the conductance of systems. This could be the connection between theoretical prediction and experimental results.  
We have used the boundary MERA framework to classify two fixed points in a simple two-wire case shown by Kane and Fisher. By keeping explicitly the scale invariance of the boundary state we can obtain current-current correlation functions that decay as a power law with either a universal or non-universal exponent and prefactor, depending on the RG fixed point reached. We can also obtain the bulk and boundary scaling dimensions that agree perfectly with the formal RG analysis. This establishes firmly the boundary MERA as a numerical method to determine the RG fixed point and the universal conductance of quantum two-wire junctions. 

The method has the advantage that it can be easily extended to study multi-wire junctions. Even in the simplest case, the Y-junction with three LL wires, not all the fixed points are fully understood by the CFT.~\cite{Oshikawa2006} We expect that boundary MERA can provide a new approach to gain insights into the properties of possible RG fixed points and their classification for more complicated multi-wire junctions,~\cite{Lal2002,Chen2002,Egger2003,Pham2003,Kazymyrenko2005,Das2008,Bellazzini2009,Aristov2010,Aristov2011,Aristov2011b} spinful LL wires,\cite{Hou2008} junctions of LL wires with different interaction strength in each wire\cite{Safi1995,Maslov1995,Aristov2012,Hou2012,Aristov2013}, and junctions of  Josephson-Junction networks.\cite{Cirillo:2011uq,Giuliano:2013fk} Potentially, the boundary MERA also provides an unbiased numerical RG method to resolve the issue about whether the conductance of Y-junction can break the single particle unitarity in the strong attractive interaction regime.~\cite{Oshikawa2006,Aristov2011b}

In addition, since we optimize the bulk scale invariant MERA independently of the boundary, the bulk results can be reused. This potentially can significantly reduce the computational costs, and can have the advantage over the DMRG method proposed in Ref.~\onlinecite{Rahmani2010}. Moreover, the scaling dimensions of the primary fields at the impurity site can be directly obtained, which can provide crucial information about the associated boundary CFT and enable further classifications of the RG fixed points.~\cite{Wong1994,Oshikawa2006,Hou2008} While the conductance of multi-wire junctions has been calculated by CFT, however, only very few numerical calculations exist in the literature to quantitatively study  and classify these results in details. In the MERA framework, none of the theoretical manipulation required in the DMRG is necessary, and a direct computation of the current-current correlation function is possible. This provides a systematic and direct numerical method to study the effects of strong electron-electron interactions in the transport properties of quantum impurity problems and molecular electronic devices.

%%%%%%%%%%%%%%%%%%%%%%%%%%%%%%%%%%%%%%%%%%%%%%%%%%
\begin{acknowledgments}
We acknowledge the inspiring discussions with G. Evenbly, G. Vidal, M. Oshikawa, and M. Cazalilla. Chang-Yu Hou acknowledges the support from DARPA-QuEST program, Packard foundation and IQIM. 
The support from NSC in Taiwan through Grants No.~100-2112-M-002-013-MY3, 100-2923-M-004-002 -MY3, 102-2112-M-002-003-MY3, 101-2112-M-007-010-MY3 as well as the support from NTU Grant No.~101R891004 are acknowledged.

\end{acknowledgments}

%%%%%%%%%%%%%%%%%%%%%%%%%%%%%%%%%%%%%%%%%%%%%%%%%%

%%%%%%%%%%%%%%%%%%%%%%%%%%%%%%%%%%%%%%%%%%%%%%%%%%
%%%%%%%%%%%%%%%%%%%%%%%%%%%%%%%%%%%%%%%%%%%%%%%%%%
%%%%%%%%%%%%%%%%%%%%%%%%%%%%%%%%%%%%%%%%%%%%%%%%%%
\appendix
\section{Scale invariant boundary MERA} \label{SEC:bdMERA}
%%%%%%%%%%%%%%%%%%%%
%%%%%%%%%%%%%%%%%%%%
\begin{figure}[tbhp]
  \centerline{\includegraphics[width=\columnwidth]{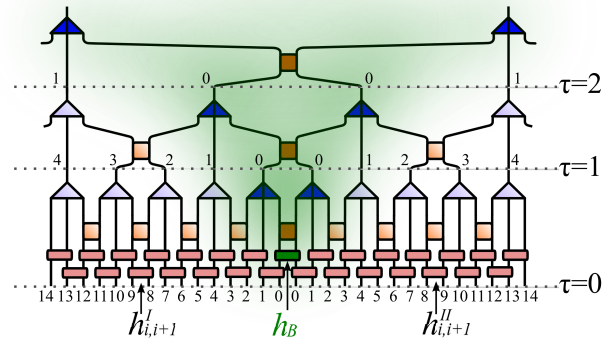}}
  \caption{(Color online) Ternary MERA for three layers with an impurity at the junction described by $h_B$. 
  The shaded green  is the causal cone for the junction. 
  The two-site Hamiltonian $h_{i,i+1}^\mu$ for layer $\tau=0$ (pink bars) is described in \eq{EQ:Hmu}.
  We also label site indices in each layer  running from zero to infinity for both wires.
The light blue triangles and the light yellow squares  represent bulk isometries and bulk disentanglers, respectively.}
  \label{FIG:Rho_W_V}
\end{figure}
%%%%%%%%%%%%%%%%%%%%
%%%%%%%%%%%%%%%%%%%%

%We note that although our implementation has  similarities to that in , we use the energy per site as our cost function for optimization 

The boundary MERA framework used in this work is based on a ternary bulk MERA of two semi-infinite wires with a junction as shown in \fig{FIG:Rho_W_V}. The shaded green area represents the casual cone associated with the junction that is described by $h_B$. It is clear from the figure that when one connects two wires with a junction, one does not need to re-optimize the MERA structure associated with the bulk part of the wires (light color tensors). The tensors in the  shaded green area, however, need to be re-optimized. To simplify the structure of the boundary MERA, we fuse tensors inside the green shaded area to form a rank-four {\em central tensors} $w_{\tau}^B$ with four external legs at each layer (\fig{fig:defectMERA}). Within the boundary MERA framework, the boundary state is characterized by these central tensors.

\begin{figure}[tb]
  \centerline{\includegraphics[width=8cm]{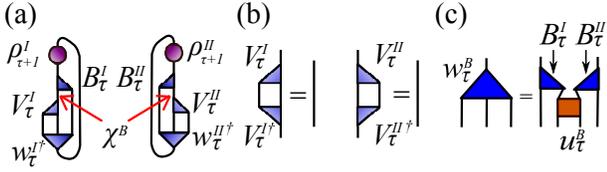}}
  \caption{ (Color online)
  (a) Graphic representation of \eq{EQ:RhoWV}. An isometry is decomposed into 
  two  tensors $V_\tau^\mu$ and $B_{\tau}^\mu$ linked in the  truncated bond dimension $\chi^B$. 
  (b) The isometric conditions of $V_\tau^\mu$ for $\mu \in \;I,II$. The lines at the right hand side of equal signs are identity matrices.
  (c) Construct a central tensor $w_\tau^B$ from three tensors $B^I_\tau$, $B^{II}_\tau$, and $u_\tau^B$. 
   }
  \label{FIG:bdMERA_trunc}
\end{figure}

We introduce both the {\em boundary truncation tensors} $V^\mu_\tau$ and the {\em boundary tensors} $B^\mu_{\tau}$ to reduce the computational cost and the memory storage during the optimization of the central tensors. 
The boundary truncation tensors $V^\mu_\tau$ allow the  bond dimensions of the central tensors to be different from the  bond dimensions of the bulk MERA tensors.
%It also allows us to have different bond dimensions on the bulk MERA tensors and the central tensors.
%For clarity we suppress the layer index $\tau$ in the following derivations. 
For simplicity, all tensors here are scale invariant, and each bond of bulk $w^\mu_\tau$  has the same bond dimension $\chi$. As shown in \fig{FIG:bdMERA_trunc}(a), both $V^\mu_\tau$ and $B^\mu_{\tau}$ are obtained by decomposing  the rank-four bulk isometry $w^\mu_\tau$  as two rank-three tensors that satisfies the equation
%%%
\begin{align} \label{EQ:RhoWV}
  \tr\{  w^{\mu\,\dagger}_\tau   V_\tau^\mu   B_\tau^\mu  \rho^\mu_{\tau+1}   \} &= 1\;,
\end{align}
%%%
where  $\rho^\mu_{\tau+1}$ is bulk one-site density matrix. When truncation is necessary, one can truncate the bond linking the $V^\mu_\tau$ and $B^\mu_{\tau}$ to some number $\chi^B \le  \chi^2$.  
%In general bond dimension $\chi^{(B)}$ can be layer dependent until the scale invariance is used to carry out same $\chi^{(B)}$ for all scale invariant layers.
We note that  $V_\tau^{\mu}$ of all possible layers satisfies the orthogonal condition $V^{\mu\dagger}_\tau \, V^\mu_\tau=\mathbf{1}$ as shown in \fig{FIG:bdMERA_trunc}(b). Conceptually, we fuse $B^I_{\tau}$, $B^{II}_{\tau}$ and the boundary disentangler $u^B_\tau$ in the green causal cone of \fig{FIG:Rho_W_V} to form the rank-four central tensor $w^B_\tau$ in \fig{fig:defectMERA} by the contraction shown in \fig{FIG:bdMERA_trunc}(c). With all the derivation above, one arrives at the MERA structure describing two-wires with a junction, as shown in  \fig{fig:defectMERA}.
We refer to  Ref.~\onlinecite{Glen13_2} for optimization details of both  boundary truncation tensors  and boundary tensors.
%%%%%%%%%%%%%%%%%%%%
%%%%%%%%%%%%%%%%%%%%
\subsection{Boundary Hamiltonian $\widetilde{H}_B$} \label{SEC:GetBdH}
%%%%%%%%%%%%%%%%%%%%%%
%%%%%%%%%%%%%%%%%%%%%%
\begin{figure}%[tbhp]
\centerline{\includegraphics[width=8cm]{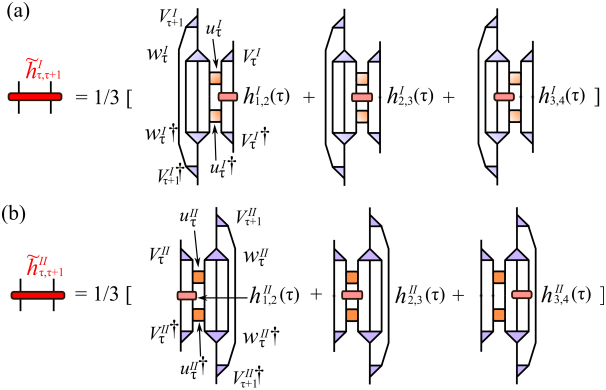}}
\caption{(Color online) Graphic representation of  $\widetilde{h}_{\tau,\tau+1}^\mu$ (a) for wire $I$ and (b) for wire $II$.}
\label{fig:ascendkalltau}
\end{figure}
%%%%%%%%%%%%%%%%%%%%%%
%%%%%%%%%%%%%%%%%%%%%%

In this section we describe how to construct the effective boundary Hamiltonian from the bare Hamiltonian of a general two-wire model:
\begin{align}
  \label{EQ:totalH}
  H &= h_{B} + H_{\rm w}\;,\\
  H_{\rm w} &= \sum_{\mu\in I, II} \sum_{i=0}^{\infty} h^\mu_{i,i+1}\; ,
\label{EQ:Hmu}
\end{align}
where $h_B$ is the on-site impurity Hamiltonian shown as the green circle in \fig{fig:defectMERA}, and $H_{\rm w}$ represents two semi-infinite Hamiltonian for wires $\mu\in I,II$. We assume that the wire Hamiltonian can be expressed as a sum of nearest-neighbor interactions in \eq{EQ:Hmu}. In particular, for the spin-1/2 XXZ model considered in this work, one has
\begin{align}
  h_{B} &= -t( S^{I\,+}_{0} S^{II\,-}_{0} + S^{I\,-}_{0} S^{II\,+}_{0}) \;, \\
  h^{I}_{i,i+1} &= J(S^{I\,X}_{i+1} S^{I\,X}_{i} +S^{I\,Y}_{i+1} S^{I\,Y}_{i}) +\lambda S^{I\,Z}_{i+1} S^{I\,Z}_{i}\;,\\
  h^{II}_{i,i+1} &= J(S^{II\,X}_{i} S^{II\,X}_{i+1}+S^{II\,Y}_{i}S^{II\,Y}_{i+1}) +\lambda S^{II\,Z}_{i}S^{II\,Z}_{i+1}\;.
\end{align}

There are two stages in constructing the effective boundary Hamiltonian:
%\begin{enumerate}
%%%%%%%%%%%
%%%%%%%%%%%
%\item 

\textit{Regrouping the bare Hamiltonian}--
  As shown in \fig{fig:defectMERA}, we regroup the bare Hamiltonian into $K^\mu_{\tau,\tau+1}$ according to \eq{EQ:K}, where $\tau$ is layer index. %Each group contains all the terms that will be ascended to become  the boundary Hamiltonian between layer $\tau$ and $\tau+1$.
Apply a sequence of average bulk ascending processes on the subset Hamiltonian $K^\mu_{\tau,\tau+1}$ until  layer $\tau$ is reached, 
\begin{align}
h^\mu_{i,i+1}(\tau) &= \overline{A}_{\rm bulk} [h^\mu_{j,j+1}(\tau-1)]\;,
\label{eq:bulkreachtau}
\end{align} 
where  $h^\mu_{i,i+1}(\tau) $ is the two-site bulk Hamiltonian in layer $\tau$, and $\overline{A}_{\rm bulk}$ is the bulk average ascending superoperator in the MERA framework.~\cite{MERAalgorithm}
%In addition, once the bulk MERA is performed with translational symmetry, the two-site Hamiltonian $h^\mu_{i,i+1}(\tau)$ should be the same for different sites in the same layer.
In addition, if we consider a translational invariant bulk MERA, within the same layer $\tau$, $h^\mu_{i,i+1}(\tau)$ remains the same for different sites due to the translational invariance.
%%%%%%%%%%%
%%%%%%%%%%%

%\item 
\textit{Performing the inhomogeneous ascending operation}--
  The two-site boundary Hamiltonian $\widetilde{h}_{\tau,\tau+1}^\mu$ is obtained by an inhomogeneous coarse-graining of bulk two-site Hamiltonians $h^\mu_{i,i+1}(\tau)$ in layer $\tau$. By applying bulk ascending process on $K_{\tau,\tau+1}^\mu$, we obtain $h^\mu_{i,i+1}(\tau)$, and 
  %After apply the bulk ascending processes on $K_{\tau,\tau+1}$, the two-site Hamiltonian $h^\mu_{i,i+1}(\tau)$ in layer $\tau$ is obtained.
we employ the inhomogeneous boundary coarse-graining with a scaling factor $1/3$ (\fig{fig:ascendkalltau}),
\begin{align}
\widetilde{h}_{\tau,\tau+1}^\mu &= \overline{A}_{\rm bd}^{\mu} [h^\mu_{1,2}(\tau)+h^\mu_{2,3}(\tau)+h^\mu_{3,4}(\tau)]\;.
\label{EQ:Bd_ascending}
\end{align} 

\begin{figure}%[tbhp]
  \centerline{\includegraphics[width=1in]{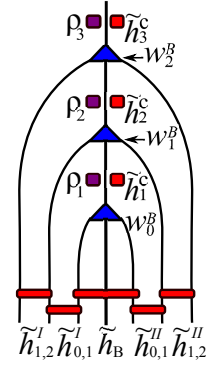}}
  \caption{(Color online)%The tensor network of the boundary MERA for three layers.
The red bars represent the effective boundary Hamiltonian consisting of $\widetilde{h}_{\tau,\tau+1}$ and  $\widetilde{h}_B$.
$w_\tau^B$ are  the central tensors indicated by blue triangles.
$\rho_\tau$  and $\widetilde{h}^{\rm c}_\tau$ are  central density matrices and central Hamiltonian, respectively. 
%the purple square.
  }
  \label{FIG:bdMERA}
\end{figure}

Once the boundary Hamiltonian is obtained, we can forget about the bulk tensors 
and concentrate on the optimization of the central tensors $w^B_\tau$ with the effective boundary Hamiltonian $\widetilde{H}$.
Therefore, the  tensor network in the boundary MERA is simplified (\fig{FIG:bdMERA}), and we perform optimization to obtain $w_\tau^B$. 
%Moreover,  we employ the same base concept of optimization in the MERA to obtain $w_\tau^B$. 
The central density matrix $\rho_\tau$ and the central Hamiltonian $\widetilde{h}^{\rm c}_\tau$ are fundamental building blocks during the updates, and they can be descended and ascended using descending and ascending superoperators described in the following section. %\App{SEC:AscendDescend}.
%We note that the effective Hamiltonian has a similar structure as the Wilson chain in the Kondo impurity problem.
%For scale invariant layers $\tau\ge \tau_s$, the two-site boundary Hamiltonian $ \sum_{\mu\in I,II} \widetilde{h}_{\tau,\tau+1}^\mu$ decays with a factor $1/3$ as one increases $\tau$.
% that is in the scaling invariant layers of the bulk MERA,
%The underlying physics can also be understood from the Wilson's NRG point of view. 
%In general, through the effective Hamiltonian in \eq{EQ:EffecH} this impurity problem 
%can be treated by any numerical method such as variational methods.
%Now the boundary MERA can be represented as the tensor network shown in \fig{FIG:bdMERA}.
%Similar to the bulk scale invariant MERA, the central tensors consist of central tensors  $\tau_1, \tau_2, \cdots, \tau_n$ for buffer layers and a single $\tau_s$ for all the  scale invariant layers.

%%%%%%%%%%
%%%%%%%%%%

%%%%%%%%%%%%%%%%%%%%%%%%%%%%%%%%%%%%%%%%
%%%%%%%%%%%%%%%%%%%%%%%%%%%%%%%%%%%%%%%%
\subsection{Central ascending and descending processes} \label{SEC:AscendDescend}
%%%%%%%%%%%%%
%%%%%%%%%%%%%
\begin{figure}%[tbhp]
\centerline{\includegraphics[width=3in]{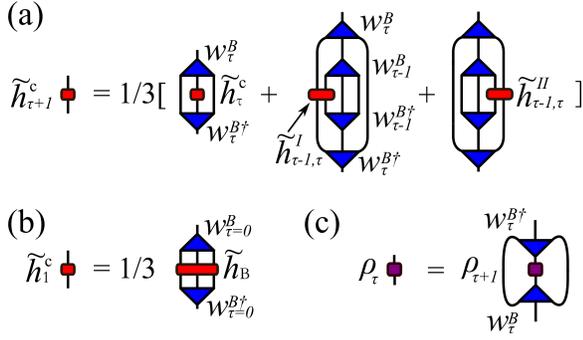}}
\caption{(Color online) (a) The average central ascending process for $\widetilde{h}^{\rm c}_{\tau+1}$.
(b)  $\widetilde{h}^{\rm c}_{1}$ is carried out using  the ascending operation of  the effective Hamiltonian $\widetilde{h}_B$, described in \fig{fig:ascend_k}~(a).
(c) The descending superoperator $D^{\rm c}$ composes of $w^B_\tau$  and $w_\tau^{B\dagger}$, and it acts on $\rho_{\tau+1}$ to obtain the central density matrix $\rho_\tau$.}
\label{FIG:Ascending}
\end{figure}
%%%%%%%%%%%%%
%%%%%%%%%%%%%

Similar to the bulk MERA, an operator that lives on the effective boundary lattice can be RG-transformed to the next or  previous layer via central  ascending or descending superoperators. In this section we describe how to construct the central ascending and  descending superoperators. Typically one use the ascending superoperator to ascend the Hamiltonian and use the descending superoperator to descend the density matrix.

First, the central Hamiltonian $\widetilde{h}^{\rm c}_{\tau+1}$ for $\tau \ge 1$ can be obtained from the lower layer  using the central ascending superoperator (\fig{FIG:Ascending}~(a) ), 
\begin{align}
  \widetilde{h}^{\rm c}_{\tau+1} &= \overline{A}(\widetilde{h}_{\tau}^{\rm c}, \widetilde{h}_{\tau-1,\tau}^{I},\widetilde{h}_{\tau-1,\tau}^{II}) \;.
\label{EQ:HC}
\end{align}
%which is pictorially defined in for $\tau \ge 1$.
 The central ascending  for $\tau =0$ is defined slightly differently (\fig{FIG:Ascending}~(b)),
\begin{align}
  \widetilde{h}^{\rm c}_{1} &= \overline{A}_0(\widetilde{h}_{\rm B}) \;.
\label{EQ:HC0}
\end{align}

%The operator of the layer $\tau+1$ is obtained by the average of three ascending process at the layer $\tau$ as
%\begin{align}
%  o^C_{\tau+1} &= \overline{\mathbf{A}}(o_{\tau})  
%  =\frac{1}{3}\sum_{\nu\in L,C,R} \mathbf{A}^\nu (o^\nu_\tau)\;,
%  \label{EQ:central_ascdening}
%\end{align}
%where $\mathbf{A}^\nu (o^\nu_\tau)$  are defined by the contraction of tensor network as shown in \fig{FIG:Ascending}(a)-(c) respectively. The index $\nu$ is used to distinguish different tensor network structures and different $o^\nu_\tau$ is used for different contraction.
%For example, the central Hamiltonian at layer $\tau+1$ is obtained by applying central average ascending superoperator to the central Hamiltonian at layer $\tau$

Second, we show how to perform central descending superoperator on the central density matrix. In contrast to the ascending superoperators, there is only one tensor network associated with the central descending superoperator  ${D}^c$ consisting of both  $w^B_\tau$ and $w^{B\dagger}_\tau$ as shown in \fig{FIG:Ascending}~(c). The central density matrix at layer $\tau$ is then obtained by applying the central descending superoperator  $D^{\rm c}$ to the central density matrix at layer $\tau+1$ as
\begin{align}
  \rho_{\tau} &= D^{\rm c}(\rho_{\tau+1}).
\end{align}

%%%%%%%%%%%%%%%%%%%%%%%%%%%%%%%%%%%%%%%%
%%%%%%%%%%%%%%%%%%%%%%%%%%%%%%
\section{Optimization of central tensors} \label{SEC:Optimization}

\begin{figure}
\centerline{\includegraphics[width=8cm]{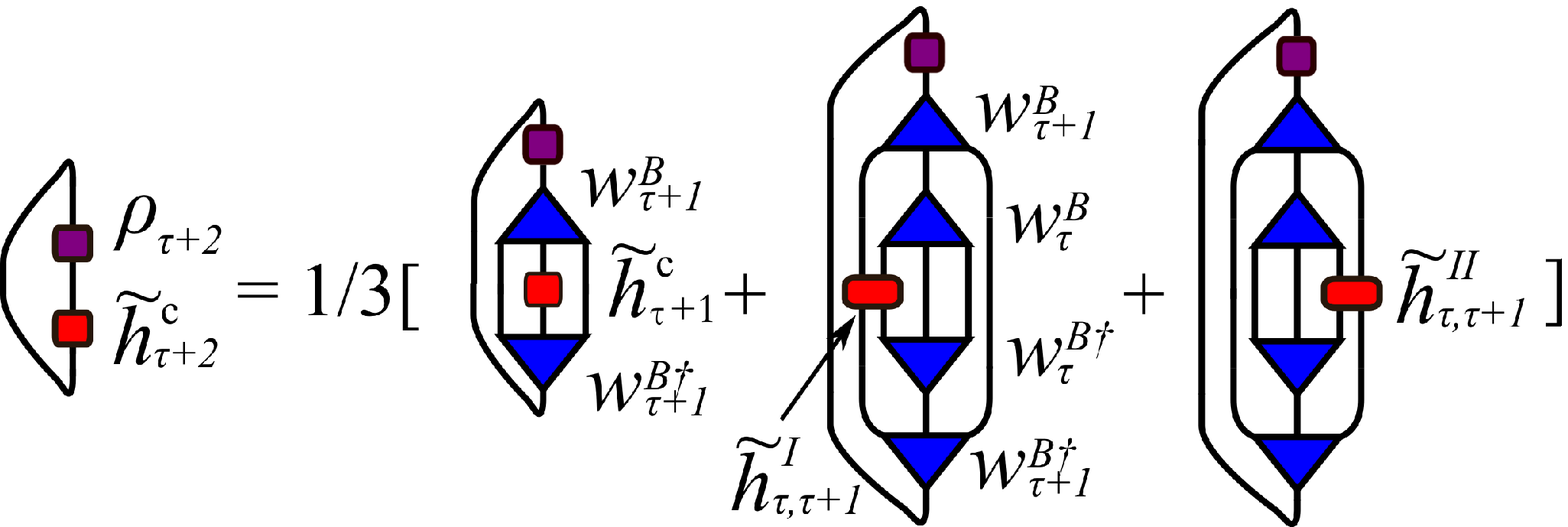}}
\caption{(Color online) Graphic representation of \eq{EQ:equilibrium} and \eq{EQ:equilibrium2} }
\label{FIG:COST_FROM}
\end{figure}

In this section we describe how to optimize the central tensor $w^B_\tau$ in  \fig{FIG:bdMERA}. We assume  several buffer layers with central tensors $w^B_\tau$, $\tau=0,1,\cdots,\tau_{{\rm bf}-1}$, before the scale invariant layers characterized by a single central tensor $w^B_s$. The optimization procedures for buffer layers and scale invariant layers are different.
% and the computational cost to update $w^B_s$ is actually cheaper due to the scale invariant properties. 
In the following we show the optimization procedure for the buffer and the scale invariant layers, respectively.

\begin{figure}%[tbhp]
  \centerline{\includegraphics[width=3in]{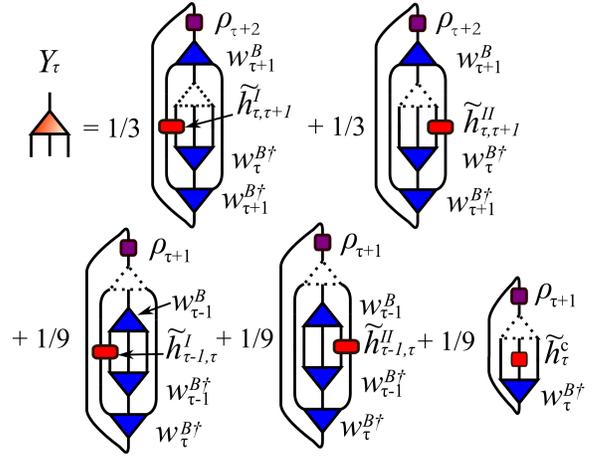}}
  \caption{ (Color online)
  The missing  triangle corresponds $w^B_\tau$, and  its corresponding  environment $Y_\tau$ 
  is defined by the sum of five tensor networks with certain weights. 
  }
  \label{FIG:Y}
\end{figure}

\begin{figure}%[tbhp]
  \centerline{\includegraphics[width=3in]{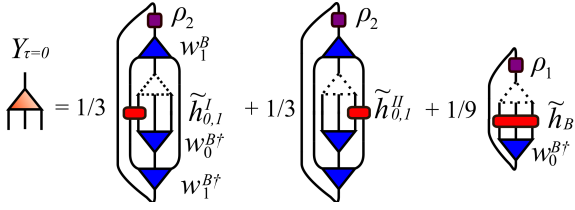}}
  \caption{ (Color online)
  The missing triangle corresponds to  $w^B_0$, and  its  environment  $Y_0$ includes three tensor networks, 
  where the effective Hamiltonian $\widetilde{h}_B $ is defined in \fig{fig:ascend_k}~(a). }
  \label{FIG:Y0}
\end{figure}

%\begin{enumerate}
%%%%%%%%%%%%%%
%\item 
\textit{Optimization in buffer layers}--
To find the optimal central tensor $w^B_\tau$, the central Hamiltonian coming from the boundary Hamiltonian including the scaling factors $1/3^\tau$ plays an important role.
We use the energy of layer $\tau+2$ as the cost function,
\begin{align}
  E_{b} &=\tr \left\{ \rho_{\tau+2} \widetilde{h}^{\rm c}_{\tau+2}\right\} \;.
  \label{EQ:equilibrium}
\end{align}  
%which consists of central tensors $ \rho_{\tau+2}$ and central Hamiltonian $ \widetilde{h}^{\rm c}_{\tau+2}$ in layer $\tau+2$.
Moreover, the central Hamiltonian $ \widetilde{h}^{\rm c}_{\tau+2}$ is obtained by the average central ascending process as shown in \eq{EQ:HC}, thus \eq{EQ:equilibrium} becomes
\begin{align}
   E_b &=\tr \left\{ \rho_{\tau+2} \overline{A}(\widetilde{h}_{\tau+1}^{\rm c}, \widetilde{h}_{\tau,\tau+1}^{I},\widetilde{h}_{\tau,\tau+1}^{II})\right\} \;,
  \label{EQ:equilibrium2}
\end{align}
which is represented graphically in \fig{FIG:COST_FROM}.
Using the same trick again, the central ascending process of $\widetilde{h}_{\tau}^{\rm c}$ replaces $\widetilde{h}_{\tau+1}^{\rm c}$ in  \eq{EQ:equilibrium2}, and the energy per site is written as, 
\begin{align}
\nonumber
  E_b &=\tr \left\{ \rho_{\tau+2} \overline{A}[
  \overline{A}(\widetilde{h}_{\tau}^{\rm c}, \widetilde{h}_{\tau-1,\tau}^{I},\widetilde{h}_{\tau-1,\tau}^{II})
  , \widetilde{h}_{\tau,\tau+1}^{I},\widetilde{h}_{\tau,\tau+1}^{II}]\right\} \;,\\
  & = \tr\left\{w^B_\tau Y_\tau \right\}\;, 
  \label{EQ:cost}
\end{align}
where $Y_\tau$  is the environment as shown in Fig.\ref{FIG:Y}. Because the environment $Y_\tau$ also contains the conjugate term $w^{B\dagger}_\tau$,  iterative process is utilized to reach the self-consistency required by \eq{EQ:cost}. Iteratively perform singular value decomposition of $Y_\tau = V \lambda U^\dagger$ to obtain  the optimal $w^B_\tau =  -UV^\dagger$. We note that the environment of the zeroth layer, $Y_0$,  has a special structure as shown in Fig.\ref{FIG:Y0}.
%%%%%
%%%%%
%%%%%
%%%%%

\begin{figure}%[tbhp]
 \centerline{\includegraphics[width=3in]{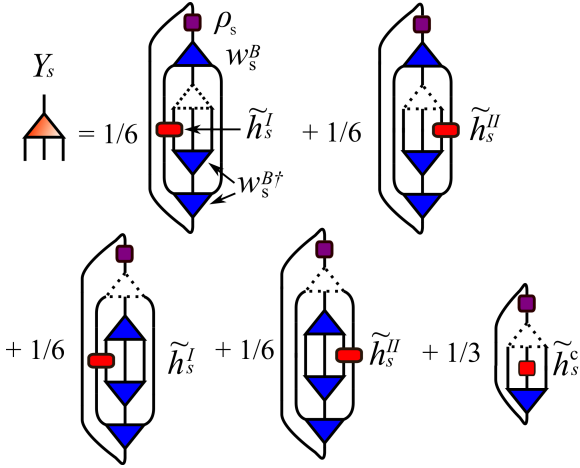}}
 \caption{ (Color online)
Both the missing and the blue triangles represent the scale invariant central tensor  $w^B_s$. 
The environment $Y_s$ is a sum of tensor networks composed of  the  scale invariant density matrix  $\rho_s$ and 
the effective scale invariant Hamiltonian  $\widetilde{h}_s^\nu$ with $\nu \in I,II,c$ defined in \eq{EQ:HS1} and  \eq{EQ:HSINF}.
 }
 \label{FIG:Ys}
\end{figure}

%%%%%%%%%%%%%%
%\item 
\textit{Optimization in scale invariant layers}--
%We next describe how to optimize the central tensor $w_s$ for all scale invariant layers. 
%To find the  optimal scale invariant central tensor $w_s$, we choose 
Similarly optimizing the central tensor of buffer layers in \eq{EQ:cost}, one can define the corresponding environment to numerically obtain the scale invariant central tensors $w_s^B$.  For the scale invariant layers, to find the optimal  $w_s^B$,   the cost function is  defined as
\begin{align}
	E_s &= 
	tr\left\{w^B_s Y_s\right\},
\label{EQ:WSYS}
\end{align}
where the corresponding environment $Y_s$  is a function of  $w_s^B$, $w_s^{B\dagger}$, $\rho_s$, and $\widetilde{h}^\nu_s$ with $\nu\in I,II,{\rm c}$. The  environment   $Y_s$ is a weighted sum of tensor networks as shown in \fig{FIG:Ys}. We note that, however, at the first scale invariant layer,  one should calculate the environment using \fig{FIG:Y} because the central tensor below is not $w^B_s$ but $w^B_{\tau_{bf-1}}$.
This means that the environment of $w^B_{s}$ is distinct from that of $w^B_\tau$ for $\tau < \tau_s+1$.
On the other hand, after the second scale invariant layer, the definition of the environment $Y_s$ in \fig{FIG:Ys} is used, since all the next layers are characterized by  the same tensor  $w^B_s$. 

%%%
%%%

%The effective scale invariant Hamiltonian $\widetilde{h}^\mu_s$ with $\mu\in I,II$ have been expounded in \eq{EQ:HS1},
%and another  effective scale invariant Hamiltonian is defined as,
We here define a scale invariant central Hamiltonian,
\begin{align}
\widetilde{h}^{\rm c}_s &=\sum_{\tau=\tau'_s}^\infty \frac{1}{3^{\tau-\tau'_s}} \widetilde{h}_\tau^{\rm c}\;,
\label{EQ:HSINF}
\end{align} 
where $\tau$ starts from the third scale invariant layer $\tau'_s = \tau_s+2$, and the construction of the central Hamiltonian $\widetilde{h}_\tau^{\rm c}$ is referred to \eq{EQ:HC}.
Moreover, from the numerical simulation perspective it is impossible to perform the infinite sum in \eq{EQ:HSINF}; 
therefore, a cut-off of  finite $L$ layers is introduced in the infinite sum. 
%However, this replacement may induce some numerical error.
%From RG point of view, 
Due to the scale invariance, the two-site Hamiltonian $\widetilde{h}_{\tau,\tau+1}^\mu$ decays quickly as a power of $1/9$ when  $\tau$  increases.~\cite{MERABCFT}
%and  indicates a certain  energy scale. 
Therefore, it is suitable to keep a finite number of $\widetilde{h}_{\tau,\tau+1}^\mu$ 
%\textbf{and the last one is in the energy scale which we are interested in (???)}.

%Here we describe a method to obtain $h_s^\mu$  in \eq{EQ:HSINF} by using the advantage of the canonical form of $V_\mu$ in \fig{FIG:Rho_W_V}(c). 
%\begin{itemize}
%\item[Step1] Apply the boundary average ascending on the bulk two-site Hamiltonian density of the $\tau$-layer, $\overline{h}^\mu_{\tau+1,\tau+2}=\overline{A}_{bd}(h_\tau^\mu)$, and the $\tau$-layer must belong to the scale invariant layers.
%\item[Step2] Utilize the bulk average ascending operator to get the next layer $h_{\tau+1}^\mu$, and go back Step1.
%\item[Step3] Sum over all two-site boundary Hamiltonian density in \eq{EQ:HSINF} with a cuff-off layer L.
%\end{itemize}
%\end{enumerate}

%%%%%%%%%%%%%%%%%%%%%%%%%%%%%%%%%%%%%%%%
\subsection{Algorithm of scale invariant boundary MERA}\label{SEC:UpdateProcess}
We briefly outline the overall update procedure for the central tensors in \fig{FIG:bdMERA}:
%and list the relevant tensors at each steps. The procedure is sketched in \fig{FIG:UpdateProcess}.

\begin{itemize}
\setlength{\itemindent}{0.2in}
  \item[Step~1.] Initialize $\widetilde{h}^{\rm c}_\tau$, $\widetilde{h}^\mu_{\tau,\tau+1}$, $\rho_\tau$, and $w^B_\tau$.
  The Hamiltonian $\widetilde{h}_{\tau,\tau+1}^I$ and $\widetilde{h}_{\tau,\tau+1}^{II}$ are obtained by the inhomogeneous coarse-graining in \eq{EQ:Bd_ascending}, 
  and $\widetilde{h}_\tau^{\rm c}$ are carried out using the central ascending processes in \eq{EQ:HC} and in \eq{EQ:HC0}. 
  For $\rho_\tau$ and $w^B_\tau$, we initialize them with the bulk tensors.  
  \item[Step~2.] Calculate the corresponding environment $Y_\tau$ of $w^B_\tau$ starting from the zeroth layer, 
  and optimize $w^B_\tau$ by minimizing the cost function in \eq{EQ:cost}. 
  Iterative  optimization is employed to acquire self-consistency for both $w^B_\tau$ and $Y_\tau$.
  \item[Step~3.] Apply the average central ascending superoperator $\overline{A}$ to obtain the  central Hamiltonian $\widetilde{h}_{\tau+1}^{\rm c}$ for the next layer.
  \item[Step~4.] Go to Step~2 for the  optimization of the next layer $(\tau+1)$ until the \textit{second} scale invariant layer is reached.
  \item[Step~5.] Apply the power method to obtain an optimal $\rho_s$. The scale invariant density matrix $\rho_s$  are the same for all the scale invariant layers.
  \item[Step~6.] Calculate the effective scale invariant Hamiltonian $\widetilde{h}_s^{I}, \widetilde{h}_s^{II}$ in \eq{EQ:HS1}, and $\widetilde{h}_s^{\rm c}$ in \eq{EQ:HSINF}.
  \item[Step~7.] Construct the scale invariant environment $Y_s$, optimize $w_s^B$ by \eq{EQ:WSYS}.
  \item[Step~8.] Apply the descending superoperator $D^{\rm c}$ on the density matrix from the top layer to the bottom and start over from Step~2.
\end{itemize}

%In the scale invariant layers, the center ascending operator becomes the scaling superoperator $\mathit{S}$,
%\begin{align}
 % \mathit{S} (\phi_\alpha) = \lambda_\alpha \phi_\alpha, ~ \lambda_\alpha = 3^{-\Delta_\alpha},
%\end{align} 
%where $\Delta_\alpha$ is the scaling dimension of the primary field.
This optimization procedure has several advantages. The most important  is the feedback between the scale invariant layers and the buffer layers. %How does the information exchange between them? 
The information of the entanglement is passed down from the scale invariant layer to the buffer layers by descending of the density matrix in Step~8. 
On the other hand, the feedback from the buffer layer to the scale invariant layer is achieved through $\widetilde{h}^{\rm c}_s$  by from the ascending of the boundary Hamiltonian and the central Hamiltonian.
%$\widetilde{h}_{\tau,\tau+1}^\mu$ and $\widetilde{h}_\tau^{\rm c}$. 
And when we optimize the central tensor $w^B_s$, we need to calculate $Y_s$ which contains the information of the effective scale invariant Hamiltonian.
This optimization method is two-way feedback such that the RG flow can more quickly reach the fixed point.
%%%%%%%%%%%
%\begin{figure}[tbhp]
%\centerline{\includegraphics[width=2in]{UpdateProcess3}}
%\caption{This is a sketch to explain the boundary MERA optimization algorithm for  acquiring the optimal central tensor  of both the buffer layers and the scale invariant layers, and we take two buffer layers as an example.}
%\label{FIG:UpdateProcess}
%\end{figure}
%%%%%%%%%%%

\section{Correlation function with perfect transmission} \label{SEC:correlation}

With the attractive electron-electron interactions, i.e., Luttinger parameter $g>1$, the presence of a single impurity is renormalized to the situation as if the impurity is in absence.~\cite{Kane1992} Hence, all correction functions are the same as an infinite Luttinger liquid wire. In this Appendix, we will focus on the equal time current-current correlation function $\langle \mathcal{T}_{\tau}J^{\mu}(x)J^{\mu'}(x)\rangle$ for $\mu\neq \mu'$ corresponding to different wires.

From Eq.~\eqref{eq:cc-function-chiral-decomposed}, we can decompose this correlation function by chiral currents as 
\begin{multline}
\langle  J^\mu(x) J^{\mu'}(x)\rangle = - v^2 \left( \langle \rho^{\mu}_{out}(x) \rho^{\mu'}_{in}(x)\rangle + \langle \rho^{\mu}_{in}(x) \rho^{\mu'}_{out}(x) \rangle \right).
\end{multline}
Here, we omit the $\mathcal{T}_\tau$ symbol. As two LL wires connected by a weak link (impurity) behave the same as a single infinite LL wires for $g>1$, the chiral current correlation functions between two wires are given by
\begin{equation}
\langle \rho^{1}_{out}(x) \rho^{2}_{in}(x)\rangle= \langle \rho^{1}_{in}(x) \rho^{2}_{out}(x)= \frac{g}{4\pi^2}\frac{1}{(2x)^2}.
\end{equation}
The normalization of the correlation function is followed by Eq.~\eqref{eq:Hw-continuum} and is consistent with Ref.~\onlinecite{Rahmani2012} The physical current-current correlation function is then given by
\begin{equation}
\langle  J^\mu(x) J^{\mu'}(x)\rangle = - \frac{v^2 g}{8\pi^2}\frac{1}{x^2} .
\end{equation}
This gives the exponent $\alpha=2$ and coefficient $A=\frac{v^2 g}{8\pi^2}$ which we will benchmark against with our numerics.

%%%%%%%%%%%%%%%%%%%%%%%%%%%%%%%%%%%%%%%%%%%%%%%%%%
\bibliography{Ref_mera}

\newcommand{\npb}{Nucl. Phys. B}\newcommand{\adv}{Adv.
  Phys.}\newcommand{\RMP}{Rev. Mod. Phys.}\newcommand{\PRB}{Phys. Rev.
  B}\newcommand{\PR}{Phys. Rev.}\newcommand{\PRL}{Phys. Rev.
  Lett.}\newcommand{\plb}{Phys. Lett.
  B}\newcommand{\jstat}{JSTAT}\newcommand{\JPC}{J. Phys,
  C}\newcommand{\njp}{New Journal of Physics}
\begin{thebibliography}{64}%
\makeatletter
\providecommand \@ifxundefined [1]{%
 \@ifx{#1\undefined}
}%
\providecommand \@ifnum [1]{%
 \ifnum #1\expandafter \@firstoftwo
 \else \expandafter \@secondoftwo
 \fi
}%
\providecommand \@ifx [1]{%
 \ifx #1\expandafter \@firstoftwo
 \else \expandafter \@secondoftwo
 \fi
}%
\providecommand \natexlab [1]{#1}%
\providecommand \enquote  [1]{``#1''}%
\providecommand \bibnamefont  [1]{#1}%
\providecommand \bibfnamefont [1]{#1}%
\providecommand \citenamefont [1]{#1}%
\providecommand \href@noop [0]{\@secondoftwo}%
\providecommand \href [0]{\begingroup \@sanitize@url \@href}%
\providecommand \@href[1]{\@@startlink{#1}\@@href}%
\providecommand \@@href[1]{\endgroup#1\@@endlink}%
\providecommand \@sanitize@url [0]{\catcode `\\12\catcode `\$12\catcode
  `\&12\catcode `\#12\catcode `\^12\catcode `\_12\catcode `\%12\relax}%
\providecommand \@@startlink[1]{}%
\providecommand \@@endlink[0]{}%
\providecommand \url  [0]{\begingroup\@sanitize@url \@url }%
\providecommand \@url [1]{\endgroup\@href {#1}{\urlprefix }}%
\providecommand \urlprefix  [0]{URL }%
\providecommand \Eprint [0]{\href }%
\providecommand \doibase [0]{http://dx.doi.org/}%
\providecommand \selectlanguage [0]{\@gobble}%
\providecommand \bibinfo  [0]{\@secondoftwo}%
\providecommand \bibfield  [0]{\@secondoftwo}%
\providecommand \translation [1]{[#1]}%
\providecommand \BibitemOpen [0]{}%
\providecommand \bibitemStop [0]{}%
\providecommand \bibitemNoStop [0]{.\EOS\space}%
\providecommand \EOS [0]{\spacefactor3000\relax}%
\providecommand \BibitemShut  [1]{\csname bibitem#1\endcsname}%
\let\auto@bib@innerbib\@empty
%</preamble>
\bibitem [{\citenamefont {Joachim}\ and\ \citenamefont
  {Ratner}(2005)}]{Joachim:2005fk}%
  \BibitemOpen
  \bibfield  {author} {\bibinfo {author} {\bibfnamefont {C.}~\bibnamefont
  {Joachim}}\ and\ \bibinfo {author} {\bibfnamefont {M.~A.}\ \bibnamefont
  {Ratner}},\ }\href@noop {} {\bibfield  {journal} {\bibinfo  {journal} {PNAS}\
  }\textbf {\bibinfo {volume} {102}},\ \bibinfo {pages} {8801} (\bibinfo {year}
  {2005})}\BibitemShut {NoStop}%
\bibitem [{\citenamefont {Ohshiro}\ \emph {et~al.}(2012)\citenamefont
  {Ohshiro}, \citenamefont {Matsubara}, \citenamefont {Tsutsui}, \citenamefont
  {Taniguchi},\ and\ \citenamefont {Kawai}}]{dnameasure:2012}%
  \BibitemOpen
  \bibfield  {author} {\bibinfo {author} {\bibfnamefont {T.}~\bibnamefont
  {Ohshiro}}, \bibinfo {author} {\bibfnamefont {K.}~\bibnamefont {Matsubara}},
  \bibinfo {author} {\bibfnamefont {M.}~\bibnamefont {Tsutsui}}, \bibinfo
  {author} {\bibfnamefont {M.~F.~M.}\ \bibnamefont {Taniguchi}}, \ and\
  \bibinfo {author} {\bibfnamefont {T.}~\bibnamefont {Kawai}},\ }\href
  {\doibase 10.1038/srep00501} {\bibfield  {journal} {\bibinfo  {journal}
  {Scientific Reports}\ }\textbf {\bibinfo {volume} {2}} (\bibinfo {year}
  {2012}),\ 10.1038/srep00501}\BibitemShut {NoStop}%
\bibitem [{\citenamefont {Laroche}\ \emph {et~al.}(2014)\citenamefont
  {Laroche}, \citenamefont {Gervais}, \citenamefont {Lilly},\ and\
  \citenamefont {Reno}}]{Laroche:2014qf}%
  \BibitemOpen
  \bibfield  {author} {\bibinfo {author} {\bibfnamefont {D.}~\bibnamefont
  {Laroche}}, \bibinfo {author} {\bibfnamefont {G.}~\bibnamefont {Gervais}},
  \bibinfo {author} {\bibfnamefont {M.~P.}\ \bibnamefont {Lilly}}, \ and\
  \bibinfo {author} {\bibfnamefont {J.~L.}\ \bibnamefont {Reno}},\ }\href
  {\doibase 10.1126/science.1244152} {\bibfield  {journal} {\bibinfo  {journal}
  {Science}\ } (\bibinfo {year} {2014}),\ 10.1126/science.1244152}\BibitemShut
  {NoStop}%
\bibitem [{\citenamefont {Ishii}\ \emph {et~al.}(2003)\citenamefont {Ishii},
  \citenamefont {Kataura}, \citenamefont {Shiozawa}, \citenamefont {Yoshioka},
  \citenamefont {Otsubo}, \citenamefont {Takayama}, \citenamefont {Miyahara},
  \citenamefont {Suzuki}, \citenamefont {Achiba}, \citenamefont {Nakatake},
  \citenamefont {Narimura}, \citenamefont {Higashiguchi}, \citenamefont
  {Shimada}, \citenamefont {Namatame},\ and\ \citenamefont
  {Taniguchi}}]{carbontube:2003}%
  \BibitemOpen
  \bibfield  {author} {\bibinfo {author} {\bibfnamefont {H.}~\bibnamefont
  {Ishii}}, \bibinfo {author} {\bibfnamefont {H.}~\bibnamefont {Kataura}},
  \bibinfo {author} {\bibfnamefont {H.}~\bibnamefont {Shiozawa}}, \bibinfo
  {author} {\bibfnamefont {H.}~\bibnamefont {Yoshioka}}, \bibinfo {author}
  {\bibfnamefont {H.}~\bibnamefont {Otsubo}}, \bibinfo {author} {\bibfnamefont
  {Y.}~\bibnamefont {Takayama}}, \bibinfo {author} {\bibfnamefont
  {T.}~\bibnamefont {Miyahara}}, \bibinfo {author} {\bibfnamefont
  {S.}~\bibnamefont {Suzuki}}, \bibinfo {author} {\bibfnamefont
  {Y.}~\bibnamefont {Achiba}}, \bibinfo {author} {\bibfnamefont
  {M.}~\bibnamefont {Nakatake}}, \bibinfo {author} {\bibfnamefont
  {T.}~\bibnamefont {Narimura}}, \bibinfo {author} {\bibfnamefont
  {M.}~\bibnamefont {Higashiguchi}}, \bibinfo {author} {\bibfnamefont
  {K.}~\bibnamefont {Shimada}}, \bibinfo {author} {\bibfnamefont
  {H.}~\bibnamefont {Namatame}}, \ and\ \bibinfo {author} {\bibfnamefont
  {M.}~\bibnamefont {Taniguchi}},\ }\href@noop {} {\bibfield  {journal}
  {\bibinfo  {journal} {Nature}\ }\textbf {\bibinfo {volume} {426}},\ \bibinfo
  {pages} {540} (\bibinfo {year} {2003})}\BibitemShut {NoStop}%
\bibitem [{\citenamefont {Bockrath}\ \emph {et~al.}(1999)\citenamefont
  {Bockrath}, \citenamefont {Cobden}, \citenamefont {Lu}, \citenamefont
  {Rinzler}, \citenamefont {Smalley}, \citenamefont {Balents},\ and\
  \citenamefont {McEuen}}]{carbonnanotube:nature1999}%
  \BibitemOpen
  \bibfield  {author} {\bibinfo {author} {\bibfnamefont {M.}~\bibnamefont
  {Bockrath}}, \bibinfo {author} {\bibfnamefont {D.~H.}\ \bibnamefont
  {Cobden}}, \bibinfo {author} {\bibfnamefont {J.}~\bibnamefont {Lu}}, \bibinfo
  {author} {\bibfnamefont {A.~G.}\ \bibnamefont {Rinzler}}, \bibinfo {author}
  {\bibfnamefont {R.~E.}\ \bibnamefont {Smalley}}, \bibinfo {author}
  {\bibfnamefont {L.}~\bibnamefont {Balents}}, \ and\ \bibinfo {author}
  {\bibfnamefont {P.~L.}\ \bibnamefont {McEuen}},\ }\href {\doibase
  10.1038/17569} {\bibfield  {journal} {\bibinfo  {journal} {Nature}\ }\textbf
  {\bibinfo {volume} {397}},\ \bibinfo {pages} {598} (\bibinfo {year}
  {1999})}\BibitemShut {NoStop}%
\bibitem [{\citenamefont {Yao}\ \emph {et~al.}(1999)\citenamefont {Yao},
  \citenamefont {Postma}, \citenamefont {Balents},\ and\ \citenamefont
  {Dekker}}]{carbontube1999}%
  \BibitemOpen
  \bibfield  {author} {\bibinfo {author} {\bibfnamefont {Z.}~\bibnamefont
  {Yao}}, \bibinfo {author} {\bibfnamefont {H.~W.~C.}\ \bibnamefont {Postma}},
  \bibinfo {author} {\bibfnamefont {L.}~\bibnamefont {Balents}}, \ and\
  \bibinfo {author} {\bibfnamefont {C.}~\bibnamefont {Dekker}},\ }\href
  {\doibase 10.1038/46241} {\bibfield  {journal} {\bibinfo  {journal} {Nature}\
  }\textbf {\bibinfo {volume} {402}},\ \bibinfo {pages} {273} (\bibinfo {year}
  {1999})}\BibitemShut {NoStop}%
\bibitem [{\citenamefont {Kim}\ \emph {et~al.}(2007)\citenamefont {Kim},
  \citenamefont {Recher}, \citenamefont {Oliver}, \citenamefont {Yamamoto},
  \citenamefont {Kong},\ and\ \citenamefont {Dai}}]{PhysRevLett.99.036802}%
  \BibitemOpen
  \bibfield  {author} {\bibinfo {author} {\bibfnamefont {N.~Y.}\ \bibnamefont
  {Kim}}, \bibinfo {author} {\bibfnamefont {P.}~\bibnamefont {Recher}},
  \bibinfo {author} {\bibfnamefont {W.~D.}\ \bibnamefont {Oliver}}, \bibinfo
  {author} {\bibfnamefont {Y.}~\bibnamefont {Yamamoto}}, \bibinfo {author}
  {\bibfnamefont {J.}~\bibnamefont {Kong}}, \ and\ \bibinfo {author}
  {\bibfnamefont {H.}~\bibnamefont {Dai}},\ }\href {\doibase
  10.1103/PhysRevLett.99.036802} {\bibfield  {journal} {\bibinfo  {journal}
  {Phys. Rev. Lett.}\ }\textbf {\bibinfo {volume} {99}},\ \bibinfo {pages}
  {036802} (\bibinfo {year} {2007})}\BibitemShut {NoStop}%
\bibitem [{\citenamefont {Postma}\ \emph {et~al.}(2000)\citenamefont {Postma},
  \citenamefont {de~Jonge}, \citenamefont {Yao},\ and\ \citenamefont
  {Dekker}}]{PhysRevB.62.R10653}%
  \BibitemOpen
  \bibfield  {author} {\bibinfo {author} {\bibfnamefont {H.~W.~C.}\
  \bibnamefont {Postma}}, \bibinfo {author} {\bibfnamefont {M.}~\bibnamefont
  {de~Jonge}}, \bibinfo {author} {\bibfnamefont {Z.}~\bibnamefont {Yao}}, \
  and\ \bibinfo {author} {\bibfnamefont {C.}~\bibnamefont {Dekker}},\ }\href
  {\doibase 10.1103/PhysRevB.62.R10653} {\bibfield  {journal} {\bibinfo
  {journal} {Phys. Rev. B}\ }\textbf {\bibinfo {volume} {62}},\ \bibinfo
  {pages} {R10653} (\bibinfo {year} {2000})}\BibitemShut {NoStop}%
\bibitem [{\citenamefont {Cardy}(2010)}]{Cardy2010}%
  \BibitemOpen
  \bibfield  {author} {\bibinfo {author} {\bibfnamefont {J.~L.}\ \bibnamefont
  {Cardy}},\ }\href@noop {} {\emph {\bibinfo {title} {Conformal field theory
  and statistical mechanics, in J. Jacobsen et al. (eds), Exact Methods in
  Low-Dimensional Statistical Physics and Quantum Computing}}}\ (\bibinfo
  {publisher} {Oxford University Press, Oxford},\ \bibinfo {year}
  {2010})\BibitemShut {NoStop}%
\bibitem [{\citenamefont {Affleck}(2010)}]{Affleck2010}%
  \BibitemOpen
  \bibfield  {author} {\bibinfo {author} {\bibfnamefont {I.}~\bibnamefont
  {Affleck}},\ }\href@noop {} {\emph {\bibinfo {title} {Quantum impurity
  problems in condensed matter physics, in J. Jacobsen et al. (eds), Exact
  Methods in Low-Dimensional Statistical Physics and Quantum Computing}}}\
  (\bibinfo  {publisher} {Oxford University Press, Oxford},\ \bibinfo {year}
  {2010})\BibitemShut {NoStop}%
\bibitem [{\citenamefont {Luttinger}(1963)}]{Luttinger63}%
  \BibitemOpen
  \bibfield  {author} {\bibinfo {author} {\bibfnamefont {J.~M.}\ \bibnamefont
  {Luttinger}},\ }\href@noop {} {\bibfield  {journal} {\bibinfo  {journal} {J.
  Math. Phys.}\ }\textbf {\bibinfo {volume} {4}},\ \bibinfo {pages} {1154}
  (\bibinfo {year} {1963})}\BibitemShut {NoStop}%
\bibitem [{\citenamefont {Kane}\ and\ \citenamefont
  {Fisher}(1992{\natexlab{a}})}]{Kane1992}%
  \BibitemOpen
  \bibfield  {author} {\bibinfo {author} {\bibfnamefont {C.~L.}\ \bibnamefont
  {Kane}}\ and\ \bibinfo {author} {\bibfnamefont {M.~P.~A.}\ \bibnamefont
  {Fisher}},\ }\href {\doibase 10.1103/PhysRevLett.68.1220} {\bibfield
  {journal} {\bibinfo  {journal} {Phys. Rev. Lett.}\ }\textbf {\bibinfo
  {volume} {68}},\ \bibinfo {pages} {1220} (\bibinfo {year}
  {1992}{\natexlab{a}})}\BibitemShut {NoStop}%
\bibitem [{\citenamefont {Kane}\ and\ \citenamefont
  {Fisher}(1992{\natexlab{b}})}]{Kane1992b}%
  \BibitemOpen
  \bibfield  {author} {\bibinfo {author} {\bibfnamefont {C.~L.}\ \bibnamefont
  {Kane}}\ and\ \bibinfo {author} {\bibfnamefont {M.~P.~A.}\ \bibnamefont
  {Fisher}},\ }\href@noop {} {\bibfield  {journal} {\bibinfo  {journal} {\PRB}\
  }\textbf {\bibinfo {volume} {46}},\ \bibinfo {pages} {15233} (\bibinfo {year}
  {1992}{\natexlab{b}})}\BibitemShut {NoStop}%
\bibitem [{\citenamefont {Furusaki}\ and\ \citenamefont
  {Nagaosa}(1993)}]{Furusaki1993}%
  \BibitemOpen
  \bibfield  {author} {\bibinfo {author} {\bibfnamefont {A.}~\bibnamefont
  {Furusaki}}\ and\ \bibinfo {author} {\bibfnamefont {N.}~\bibnamefont
  {Nagaosa}},\ }\href@noop {} {\bibfield  {journal} {\bibinfo  {journal}
  {\PRB}\ }\textbf {\bibinfo {volume} {47}},\ \bibinfo {pages} {4631} (\bibinfo
  {year} {1993})}\BibitemShut {NoStop}%
\bibitem [{\citenamefont {Wong}\ and\ \citenamefont
  {Affleck}(1994)}]{Wong1994}%
  \BibitemOpen
  \bibfield  {author} {\bibinfo {author} {\bibfnamefont {E.}~\bibnamefont
  {Wong}}\ and\ \bibinfo {author} {\bibfnamefont {I.}~\bibnamefont {Affleck}},\
  }\href@noop {} {\bibfield  {journal} {\bibinfo  {journal} {\npb}\ }\textbf
  {\bibinfo {volume} {417}},\ \bibinfo {pages} {403} (\bibinfo {year}
  {1994})}\BibitemShut {NoStop}%
\bibitem [{\citenamefont {Matveev}\ \emph {et~al.}(1993)\citenamefont
  {Matveev}, \citenamefont {Yue},\ and\ \citenamefont {Glazman}}]{Matveev1993}%
  \BibitemOpen
  \bibfield  {author} {\bibinfo {author} {\bibfnamefont {K.~A.}\ \bibnamefont
  {Matveev}}, \bibinfo {author} {\bibfnamefont {D.}~\bibnamefont {Yue}}, \ and\
  \bibinfo {author} {\bibfnamefont {L.~I.}\ \bibnamefont {Glazman}},\ }\href
  {\doibase 10.1103/PhysRevLett.71.3351} {\bibfield  {journal} {\bibinfo
  {journal} {Phys. Rev. Lett.}\ }\textbf {\bibinfo {volume} {71}},\ \bibinfo
  {pages} {3351} (\bibinfo {year} {1993})}\BibitemShut {NoStop}%
\bibitem [{\citenamefont {Affleck}\ and\ \citenamefont
  {Ludwig}(1991)}]{Affleck1991}%
  \BibitemOpen
  \bibfield  {author} {\bibinfo {author} {\bibfnamefont {I.}~\bibnamefont
  {Affleck}}\ and\ \bibinfo {author} {\bibfnamefont {A.}~\bibnamefont
  {Ludwig}},\ }\href@noop {} {\bibfield  {journal} {\bibinfo  {journal} {\npb}\
  }\textbf {\bibinfo {volume} {352}},\ \bibinfo {pages} {849} (\bibinfo {year}
  {1991})}\BibitemShut {NoStop}%
\bibitem [{\citenamefont {Nayak}\ \emph {et~al.}(1999)\citenamefont {Nayak},
  \citenamefont {Fisher}, \citenamefont {Ludwig},\ and\ \citenamefont
  {Lin}}]{Nayak1999}%
  \BibitemOpen
  \bibfield  {author} {\bibinfo {author} {\bibfnamefont {C.}~\bibnamefont
  {Nayak}}, \bibinfo {author} {\bibfnamefont {M.~P.~A.}\ \bibnamefont
  {Fisher}}, \bibinfo {author} {\bibfnamefont {A.~W.~W.}\ \bibnamefont
  {Ludwig}}, \ and\ \bibinfo {author} {\bibfnamefont {H.~H.}\ \bibnamefont
  {Lin}},\ }\href@noop {} {\bibfield  {journal} {\bibinfo  {journal} {\PRB}\
  }\textbf {\bibinfo {volume} {59}},\ \bibinfo {pages} {15694} (\bibinfo {year}
  {1999})}\BibitemShut {NoStop}%
\bibitem [{\citenamefont {Oshikawa}\ \emph {et~al.}(2006)\citenamefont
  {Oshikawa}, \citenamefont {Chamon},\ and\ \citenamefont
  {Affleck}}]{Oshikawa2006}%
  \BibitemOpen
  \bibfield  {author} {\bibinfo {author} {\bibfnamefont {M.}~\bibnamefont
  {Oshikawa}}, \bibinfo {author} {\bibfnamefont {C.}~\bibnamefont {Chamon}}, \
  and\ \bibinfo {author} {\bibfnamefont {I.}~\bibnamefont {Affleck}},\
  }\href@noop {} {\bibfield  {journal} {\bibinfo  {journal} {\jstat}\ }\textbf
  {\bibinfo {volume} {2006}},\ \bibinfo {pages} {P02008} (\bibinfo {year}
  {2006})}\BibitemShut {NoStop}%
\bibitem [{\citenamefont {Andergassen}\ \emph {et~al.}(2004)\citenamefont
  {Andergassen}, \citenamefont {Enss}, \citenamefont {Meden}, \citenamefont
  {Metzner}, \citenamefont {Schollw\"ock},\ and\ \citenamefont
  {Sch\"onhammer}}]{Andergassen2004}%
  \BibitemOpen
  \bibfield  {author} {\bibinfo {author} {\bibfnamefont {S.}~\bibnamefont
  {Andergassen}}, \bibinfo {author} {\bibfnamefont {T.}~\bibnamefont {Enss}},
  \bibinfo {author} {\bibfnamefont {V.}~\bibnamefont {Meden}}, \bibinfo
  {author} {\bibfnamefont {W.}~\bibnamefont {Metzner}}, \bibinfo {author}
  {\bibfnamefont {U.}~\bibnamefont {Schollw\"ock}}, \ and\ \bibinfo {author}
  {\bibfnamefont {K.}~\bibnamefont {Sch\"onhammer}},\ }\href {\doibase
  10.1103/PhysRevB.70.075102} {\bibfield  {journal} {\bibinfo  {journal} {Phys.
  Rev. B}\ }\textbf {\bibinfo {volume} {70}},\ \bibinfo {pages} {075102}
  (\bibinfo {year} {2004})}\BibitemShut {NoStop}%
\bibitem [{\citenamefont {Hamamoto}\ \emph {et~al.}(2008)\citenamefont
  {Hamamoto}, \citenamefont {Imura},\ and\ \citenamefont
  {Kato}}]{Hamamoto2008}%
  \BibitemOpen
  \bibfield  {author} {\bibinfo {author} {\bibfnamefont {Y.}~\bibnamefont
  {Hamamoto}}, \bibinfo {author} {\bibfnamefont {K.-I.}\ \bibnamefont {Imura}},
  \ and\ \bibinfo {author} {\bibfnamefont {T.}~\bibnamefont {Kato}},\ }\href
  {\doibase 10.1103/PhysRevB.77.165402} {\bibfield  {journal} {\bibinfo
  {journal} {Phys. Rev. B}\ }\textbf {\bibinfo {volume} {77}},\ \bibinfo
  {pages} {165402} (\bibinfo {year} {2008})}\BibitemShut {NoStop}%
\bibitem [{\citenamefont {Freyn}\ and\ \citenamefont
  {Florens}(2011)}]{Freyn2011}%
  \BibitemOpen
  \bibfield  {author} {\bibinfo {author} {\bibfnamefont {A.}~\bibnamefont
  {Freyn}}\ and\ \bibinfo {author} {\bibfnamefont {S.}~\bibnamefont
  {Florens}},\ }\href {\doibase 10.1103/PhysRevLett.107.017201} {\bibfield
  {journal} {\bibinfo  {journal} {Phys. Rev. Lett.}\ }\textbf {\bibinfo
  {volume} {107}},\ \bibinfo {pages} {017201} (\bibinfo {year}
  {2011})}\BibitemShut {NoStop}%
\bibitem [{\citenamefont {Barnab\'e-Th\'eriault}\ \emph
  {et~al.}(2005{\natexlab{a}})\citenamefont {Barnab\'e-Th\'eriault},
  \citenamefont {Sedeki}, \citenamefont {Meden},\ and\ \citenamefont
  {Sch\"onhammer}}]{PhysRevB.71.205327}%
  \BibitemOpen
  \bibfield  {author} {\bibinfo {author} {\bibfnamefont {X.}~\bibnamefont
  {Barnab\'e-Th\'eriault}}, \bibinfo {author} {\bibfnamefont {A.}~\bibnamefont
  {Sedeki}}, \bibinfo {author} {\bibfnamefont {V.}~\bibnamefont {Meden}}, \
  and\ \bibinfo {author} {\bibfnamefont {K.}~\bibnamefont {Sch\"onhammer}},\
  }\href {\doibase 10.1103/PhysRevB.71.205327} {\bibfield  {journal} {\bibinfo
  {journal} {Phys. Rev. B}\ }\textbf {\bibinfo {volume} {71}},\ \bibinfo
  {pages} {205327} (\bibinfo {year} {2005}{\natexlab{a}})}\BibitemShut
  {NoStop}%
\bibitem [{\citenamefont {Barnab\'e-Th\'eriault}\ \emph
  {et~al.}(2005{\natexlab{b}})\citenamefont {Barnab\'e-Th\'eriault},
  \citenamefont {Sedeki}, \citenamefont {Meden},\ and\ \citenamefont
  {Sch\"onhammer}}]{PhysRevLett.94.136405}%
  \BibitemOpen
  \bibfield  {author} {\bibinfo {author} {\bibfnamefont {X.}~\bibnamefont
  {Barnab\'e-Th\'eriault}}, \bibinfo {author} {\bibfnamefont {A.}~\bibnamefont
  {Sedeki}}, \bibinfo {author} {\bibfnamefont {V.}~\bibnamefont {Meden}}, \
  and\ \bibinfo {author} {\bibfnamefont {K.}~\bibnamefont {Sch\"onhammer}},\
  }\href {\doibase 10.1103/PhysRevLett.94.136405} {\bibfield  {journal}
  {\bibinfo  {journal} {Phys. Rev. Lett.}\ }\textbf {\bibinfo {volume} {94}},\
  \bibinfo {pages} {136405} (\bibinfo {year} {2005}{\natexlab{b}})}\BibitemShut
  {NoStop}%
\bibitem [{\citenamefont {Rahmani}\ \emph {et~al.}(2010)\citenamefont
  {Rahmani}, \citenamefont {Hou}, \citenamefont {Feiguin}, \citenamefont
  {Chamon},\ and\ \citenamefont {Affleck}}]{Rahmani2010}%
  \BibitemOpen
  \bibfield  {author} {\bibinfo {author} {\bibfnamefont {A.}~\bibnamefont
  {Rahmani}}, \bibinfo {author} {\bibfnamefont {C.-Y.}\ \bibnamefont {Hou}},
  \bibinfo {author} {\bibfnamefont {A.}~\bibnamefont {Feiguin}}, \bibinfo
  {author} {\bibfnamefont {C.}~\bibnamefont {Chamon}}, \ and\ \bibinfo {author}
  {\bibfnamefont {I.}~\bibnamefont {Affleck}},\ }\href@noop {} {\bibfield
  {journal} {\bibinfo  {journal} {\prl}\ }\textbf {\bibinfo {volume} {105}},\
  \bibinfo {pages} {226803} (\bibinfo {year} {2010})}\BibitemShut {NoStop}%
\bibitem [{\citenamefont {Vidal}(2007)}]{ER}%
  \BibitemOpen
  \bibfield  {author} {\bibinfo {author} {\bibfnamefont {G.}~\bibnamefont
  {Vidal}},\ }\href {\doibase 10.1103/PhysRevLett.99.220405} {\bibfield
  {journal} {\bibinfo  {journal} {Phys. Rev. Lett.}\ }\textbf {\bibinfo
  {volume} {99}},\ \bibinfo {pages} {220405} (\bibinfo {year}
  {2007})}\BibitemShut {NoStop}%
\bibitem [{\citenamefont {Evenbly}\ \emph {et~al.}(2010)\citenamefont
  {Evenbly}, \citenamefont {Pfeifer}, \citenamefont {Pic\'o}, \citenamefont
  {Iblisdir}, \citenamefont {Tagliacozzo}, \citenamefont {McCulloch},\ and\
  \citenamefont {Vidal}}]{MERABCFT}%
  \BibitemOpen
  \bibfield  {author} {\bibinfo {author} {\bibfnamefont {G.}~\bibnamefont
  {Evenbly}}, \bibinfo {author} {\bibfnamefont {R.~N.~C.}\ \bibnamefont
  {Pfeifer}}, \bibinfo {author} {\bibfnamefont {V.}~\bibnamefont {Pic\'o}},
  \bibinfo {author} {\bibfnamefont {S.}~\bibnamefont {Iblisdir}}, \bibinfo
  {author} {\bibfnamefont {L.}~\bibnamefont {Tagliacozzo}}, \bibinfo {author}
  {\bibfnamefont {I.~P.}\ \bibnamefont {McCulloch}}, \ and\ \bibinfo {author}
  {\bibfnamefont {G.}~\bibnamefont {Vidal}},\ }\href {\doibase
  10.1103/PhysRevB.82.161107} {\bibfield  {journal} {\bibinfo  {journal} {Phys.
  Rev. B}\ }\textbf {\bibinfo {volume} {82}},\ \bibinfo {pages} {161107}
  (\bibinfo {year} {2010})}\BibitemShut {NoStop}%
\bibitem [{\citenamefont {White}(1992)}]{White}%
  \BibitemOpen
  \bibfield  {author} {\bibinfo {author} {\bibfnamefont {S.~R.}\ \bibnamefont
  {White}},\ }\href {\doibase 10.1103/PhysRevLett.69.2863} {\bibfield
  {journal} {\bibinfo  {journal} {Phys. Rev. Lett.}\ }\textbf {\bibinfo
  {volume} {69}},\ \bibinfo {pages} {2863} (\bibinfo {year}
  {1992})}\BibitemShut {NoStop}%
\bibitem [{\citenamefont {Schollw\"ock}(2005)}]{Schollwock:2005vn}%
  \BibitemOpen
  \bibfield  {author} {\bibinfo {author} {\bibfnamefont {U.}~\bibnamefont
  {Schollw\"ock}},\ }\href {\doibase 10.1103/RevModPhys.77.259} {\bibfield
  {journal} {\bibinfo  {journal} {Rev. Mod. Phys.}\ }\textbf {\bibinfo {volume}
  {77}},\ \bibinfo {pages} {259} (\bibinfo {year} {2005})}\BibitemShut
  {NoStop}%
\bibitem [{\citenamefont {Meden}\ and\ \citenamefont
  {Schollw\"ock}(2003)}]{Meden2003}%
  \BibitemOpen
  \bibfield  {author} {\bibinfo {author} {\bibfnamefont {V.}~\bibnamefont
  {Meden}}\ and\ \bibinfo {author} {\bibfnamefont {U.}~\bibnamefont
  {Schollw\"ock}},\ }\href {\doibase 10.1103/PhysRevB.67.193303} {\bibfield
  {journal} {\bibinfo  {journal} {Phys. Rev. B}\ }\textbf {\bibinfo {volume}
  {67}},\ \bibinfo {pages} {193303} (\bibinfo {year} {2003})}\BibitemShut
  {NoStop}%
\bibitem [{\citenamefont {Rahmani}\ \emph {et~al.}(2012)\citenamefont
  {Rahmani}, \citenamefont {Hou}, \citenamefont {Feiguin}, \citenamefont
  {Oshikawa}, \citenamefont {Chamon},\ and\ \citenamefont
  {Affleck}}]{Rahmani2012}%
  \BibitemOpen
  \bibfield  {author} {\bibinfo {author} {\bibfnamefont {A.}~\bibnamefont
  {Rahmani}}, \bibinfo {author} {\bibfnamefont {C.-Y.}\ \bibnamefont {Hou}},
  \bibinfo {author} {\bibfnamefont {A.}~\bibnamefont {Feiguin}}, \bibinfo
  {author} {\bibfnamefont {M.}~\bibnamefont {Oshikawa}}, \bibinfo {author}
  {\bibfnamefont {C.}~\bibnamefont {Chamon}}, \ and\ \bibinfo {author}
  {\bibfnamefont {I.}~\bibnamefont {Affleck}},\ }\href@noop {} {\bibfield
  {journal} {\bibinfo  {journal} {Phys. Rev. B}\ }\textbf {\bibinfo {volume}
  {85}},\ \bibinfo {pages} {045120} (\bibinfo {year} {2012})}\BibitemShut
  {NoStop}%
\bibitem [{\citenamefont {Karrasch}\ and\ \citenamefont
  {Moore}(2012)}]{Karrasch2012}%
  \BibitemOpen
  \bibfield  {author} {\bibinfo {author} {\bibfnamefont {C.}~\bibnamefont
  {Karrasch}}\ and\ \bibinfo {author} {\bibfnamefont {J.~E.}\ \bibnamefont
  {Moore}},\ }\href@noop {} {\bibfield  {journal} {\bibinfo  {journal} {Phys.
  Rev. B}\ }\textbf {\bibinfo {volume} {86}},\ \bibinfo {pages} {155156}
  (\bibinfo {year} {2012})}\BibitemShut {NoStop}%
\bibitem [{\citenamefont {Eggert}\ and\ \citenamefont
  {Affleck}(1992)}]{PhysRevB.46.10866}%
  \BibitemOpen
  \bibfield  {author} {\bibinfo {author} {\bibfnamefont {S.}~\bibnamefont
  {Eggert}}\ and\ \bibinfo {author} {\bibfnamefont {I.}~\bibnamefont
  {Affleck}},\ }\href {\doibase 10.1103/PhysRevB.46.10866} {\bibfield
  {journal} {\bibinfo  {journal} {Phys. Rev. B}\ }\textbf {\bibinfo {volume}
  {46}},\ \bibinfo {pages} {10866} (\bibinfo {year} {1992})}\BibitemShut
  {NoStop}%
\bibitem [{\citenamefont {Qin}\ \emph {et~al.}(1996)\citenamefont {Qin},
  \citenamefont {Fabrizio},\ and\ \citenamefont {Yu}}]{PhysRevB.54.R9643}%
  \BibitemOpen
  \bibfield  {author} {\bibinfo {author} {\bibfnamefont {S.}~\bibnamefont
  {Qin}}, \bibinfo {author} {\bibfnamefont {M.}~\bibnamefont {Fabrizio}}, \
  and\ \bibinfo {author} {\bibfnamefont {L.}~\bibnamefont {Yu}},\ }\href
  {\doibase 10.1103/PhysRevB.54.R9643} {\bibfield  {journal} {\bibinfo
  {journal} {Phys. Rev. B}\ }\textbf {\bibinfo {volume} {54}},\ \bibinfo
  {pages} {R9643} (\bibinfo {year} {1996})}\BibitemShut {NoStop}%
\bibitem [{\citenamefont {Enss}\ \emph {et~al.}(2005)\citenamefont {Enss},
  \citenamefont {Meden}, \citenamefont {Andergassen}, \citenamefont
  {Barnab\'e-Th\'eriault}, \citenamefont {Metzner},\ and\ \citenamefont
  {Sch\"onhammer}}]{Enss2005}%
  \BibitemOpen
  \bibfield  {author} {\bibinfo {author} {\bibfnamefont {T.}~\bibnamefont
  {Enss}}, \bibinfo {author} {\bibfnamefont {V.}~\bibnamefont {Meden}},
  \bibinfo {author} {\bibfnamefont {S.}~\bibnamefont {Andergassen}}, \bibinfo
  {author} {\bibfnamefont {X.}~\bibnamefont {Barnab\'e-Th\'eriault}}, \bibinfo
  {author} {\bibfnamefont {W.}~\bibnamefont {Metzner}}, \ and\ \bibinfo
  {author} {\bibfnamefont {K.}~\bibnamefont {Sch\"onhammer}},\ }\href {\doibase
  10.1103/PhysRevB.71.155401} {\bibfield  {journal} {\bibinfo  {journal} {Phys.
  Rev. B}\ }\textbf {\bibinfo {volume} {71}},\ \bibinfo {pages} {155401}
  (\bibinfo {year} {2005})}\BibitemShut {NoStop}%
\bibitem [{\citenamefont {Andergassen}\ \emph {et~al.}(2006)\citenamefont
  {Andergassen}, \citenamefont {Enss}, \citenamefont {Meden}, \citenamefont
  {Metzner}, \citenamefont {Schollw\"ock},\ and\ \citenamefont
  {Sch\"onhammer}}]{Andergassen2006}%
  \BibitemOpen
  \bibfield  {author} {\bibinfo {author} {\bibfnamefont {S.}~\bibnamefont
  {Andergassen}}, \bibinfo {author} {\bibfnamefont {T.}~\bibnamefont {Enss}},
  \bibinfo {author} {\bibfnamefont {V.}~\bibnamefont {Meden}}, \bibinfo
  {author} {\bibfnamefont {W.}~\bibnamefont {Metzner}}, \bibinfo {author}
  {\bibfnamefont {U.}~\bibnamefont {Schollw\"ock}}, \ and\ \bibinfo {author}
  {\bibfnamefont {K.}~\bibnamefont {Sch\"onhammer}},\ }\href {\doibase
  10.1103/PhysRevB.73.045125} {\bibfield  {journal} {\bibinfo  {journal} {Phys.
  Rev. B}\ }\textbf {\bibinfo {volume} {73}},\ \bibinfo {pages} {045125}
  (\bibinfo {year} {2006})}\BibitemShut {NoStop}%
\bibitem [{\citenamefont {Evenbly}\ and\ \citenamefont
  {Vidal}(2009)}]{MERAalgorithm}%
  \BibitemOpen
  \bibfield  {author} {\bibinfo {author} {\bibfnamefont {G.}~\bibnamefont
  {Evenbly}}\ and\ \bibinfo {author} {\bibfnamefont {G.}~\bibnamefont
  {Vidal}},\ }\href {\doibase 10.1103/PhysRevB.79.144108} {\bibfield  {journal}
  {\bibinfo  {journal} {Phys. Rev. B}\ }\textbf {\bibinfo {volume} {79}},\
  \bibinfo {pages} {144108} (\bibinfo {year} {2009})}\BibitemShut {NoStop}%
\bibitem [{\citenamefont {Pfeifer}\ \emph {et~al.}(2009)\citenamefont
  {Pfeifer}, \citenamefont {Evenbly},\ and\ \citenamefont {Vidal}}]{MERACFT}%
  \BibitemOpen
  \bibfield  {author} {\bibinfo {author} {\bibfnamefont {R.~N.~C.}\
  \bibnamefont {Pfeifer}}, \bibinfo {author} {\bibfnamefont {G.}~\bibnamefont
  {Evenbly}}, \ and\ \bibinfo {author} {\bibfnamefont {G.}~\bibnamefont
  {Vidal}},\ }\href {\doibase 10.1103/PhysRevA.79.040301} {\bibfield  {journal}
  {\bibinfo  {journal} {Phys. Rev. A}\ }\textbf {\bibinfo {volume} {79}},\
  \bibinfo {pages} {040301} (\bibinfo {year} {2009})}\BibitemShut {NoStop}%
\bibitem [{\citenamefont {Evenbly}\ and\ \citenamefont
  {Vidal}(2013{\natexlab{a}})}]{Glen13_2}%
  \BibitemOpen
  \bibfield  {author} {\bibinfo {author} {\bibfnamefont {G.}~\bibnamefont
  {Evenbly}}\ and\ \bibinfo {author} {\bibfnamefont {G.}~\bibnamefont
  {Vidal}},\ }\href {http://arxiv.org/abs/1312.0303} {} (\bibinfo {year}
  {2013}{\natexlab{a}}),\ \Eprint {http://arxiv.org/abs/1312.0303}
  {arXiv:1312.0303} \BibitemShut {NoStop}%
\bibitem [{\citenamefont {Weng}(2010)}]{WengER}%
  \BibitemOpen
  \bibfield  {author} {\bibinfo {author} {\bibfnamefont {M.~Q.}\ \bibnamefont
  {Weng}},\ }\href {http://stacks.iop.org/0295-5075/92/i=6/a=60005} {\bibfield
  {journal} {\bibinfo  {journal} {EPL (Europhysics Letters)}\ }\textbf
  {\bibinfo {volume} {92}},\ \bibinfo {pages} {60005} (\bibinfo {year}
  {2010})}\BibitemShut {NoStop}%
\bibitem [{\citenamefont {Silvi}\ \emph {et~al.}(2010)\citenamefont {Silvi},
  \citenamefont {Giovannetti}, \citenamefont {Calabrese}, \citenamefont
  {Santoro},\ and\ \citenamefont {Fazio}}]{FazioER}%
  \BibitemOpen
  \bibfield  {author} {\bibinfo {author} {\bibfnamefont {P.}~\bibnamefont
  {Silvi}}, \bibinfo {author} {\bibfnamefont {V.}~\bibnamefont {Giovannetti}},
  \bibinfo {author} {\bibfnamefont {P.}~\bibnamefont {Calabrese}}, \bibinfo
  {author} {\bibfnamefont {G.~E.}\ \bibnamefont {Santoro}}, \ and\ \bibinfo
  {author} {\bibfnamefont {R.}~\bibnamefont {Fazio}},\ }\href
  {http://stacks.iop.org/1742-5468/2010/i=03/a=L03001} {\bibfield  {journal}
  {\bibinfo  {journal} {Journal of Statistical Mechanics: Theory and
  Experiment}\ }\textbf {\bibinfo {volume} {2010}},\ \bibinfo {pages} {L03001}
  (\bibinfo {year} {2010})}\BibitemShut {NoStop}%
\bibitem [{\citenamefont {Cardy}\ and\ \citenamefont
  {Lewellen}(1991)}]{Cardy1991}%
  \BibitemOpen
  \bibfield  {author} {\bibinfo {author} {\bibfnamefont {J.~L.}\ \bibnamefont
  {Cardy}}\ and\ \bibinfo {author} {\bibfnamefont {D.~C.}\ \bibnamefont
  {Lewellen}},\ }\href@noop {} {\bibfield  {journal} {\bibinfo  {journal}
  {\plb}\ }\textbf {\bibinfo {volume} {259}},\ \bibinfo {pages} {274} (\bibinfo
  {year} {1991})}\BibitemShut {NoStop}%
\bibitem [{\citenamefont {Evenbly}\ and\ \citenamefont
  {Vidal}(2013{\natexlab{b}})}]{Glen13_1}%
  \BibitemOpen
  \bibfield  {author} {\bibinfo {author} {\bibfnamefont {G.}~\bibnamefont
  {Evenbly}}\ and\ \bibinfo {author} {\bibfnamefont {G.}~\bibnamefont
  {Vidal}},\ }\href {http://arxiv.org/abs/1307.0831} {} (\bibinfo {year}
  {2013}{\natexlab{b}}),\ \Eprint {http://arxiv.org/abs/1307.0831}
  {arXiv:1307.0831} \BibitemShut {NoStop}%
\bibitem [{\citenamefont {Avella}\ and\ \citenamefont
  {Mancini}(2013)}]{MERAbook}%
  \BibitemOpen
  \bibfield  {author} {\bibinfo {author} {\bibfnamefont {A.}~\bibnamefont
  {Avella}}\ and\ \bibinfo {author} {\bibfnamefont {F.}~\bibnamefont
  {Mancini}},\ }\href@noop {} {\emph {\bibinfo {title} {Strongly Correlated
  Systems}}},\ \bibinfo {series} {Springer Series in Solid-State Sciences},
  Vol.\ \bibinfo {volume} {176}\ (\bibinfo  {publisher} {Springer},\ \bibinfo
  {year} {2013})\BibitemShut {NoStop}%
\bibitem [{\citenamefont {Lukyanov}(1998)}]{Lukyanov1998}%
  \BibitemOpen
  \bibfield  {author} {\bibinfo {author} {\bibfnamefont {S.}~\bibnamefont
  {Lukyanov}},\ }\href@noop {} {\bibfield  {journal} {\bibinfo  {journal}
  {Nuclear Physics B}\ }\textbf {\bibinfo {volume} {522}},\ \bibinfo {pages}
  {533 } (\bibinfo {year} {1998})}\BibitemShut {NoStop}%
\bibitem [{\citenamefont {Lukyanov}\ and\ \citenamefont
  {Terras}(2003)}]{Lukyanov2003323}%
  \BibitemOpen
  \bibfield  {author} {\bibinfo {author} {\bibfnamefont {S.}~\bibnamefont
  {Lukyanov}}\ and\ \bibinfo {author} {\bibfnamefont {V.}~\bibnamefont
  {Terras}},\ }\href {\doibase http://dx.doi.org/10.1016/S0550-3213(02)01141-0}
  {\bibfield  {journal} {\bibinfo  {journal} {Nuclear Physics B}\ }\textbf
  {\bibinfo {volume} {654}},\ \bibinfo {pages} {323 } (\bibinfo {year}
  {2003})}\BibitemShut {NoStop}%
\bibitem [{\citenamefont {Lal}\ \emph {et~al.}(2002)\citenamefont {Lal},
  \citenamefont {Rao},\ and\ \citenamefont {Sen}}]{Lal2002}%
  \BibitemOpen
  \bibfield  {author} {\bibinfo {author} {\bibfnamefont {S.}~\bibnamefont
  {Lal}}, \bibinfo {author} {\bibfnamefont {S.}~\bibnamefont {Rao}}, \ and\
  \bibinfo {author} {\bibfnamefont {D.}~\bibnamefont {Sen}},\ }\href@noop {}
  {\bibfield  {journal} {\bibinfo  {journal} {\PRB}\ }\textbf {\bibinfo
  {volume} {66}},\ \bibinfo {pages} {165327} (\bibinfo {year}
  {2002})}\BibitemShut {NoStop}%
\bibitem [{\citenamefont {Chen}\ \emph {et~al.}(2002)\citenamefont {Chen},
  \citenamefont {Trauzettel},\ and\ \citenamefont {Egger}}]{Chen2002}%
  \BibitemOpen
  \bibfield  {author} {\bibinfo {author} {\bibfnamefont {S.}~\bibnamefont
  {Chen}}, \bibinfo {author} {\bibfnamefont {B.}~\bibnamefont {Trauzettel}}, \
  and\ \bibinfo {author} {\bibfnamefont {R.}~\bibnamefont {Egger}},\ }\href
  {\doibase 10.1103/PhysRevLett.89.226404} {\bibfield  {journal} {\bibinfo
  {journal} {Phys. Rev. Lett.}\ }\textbf {\bibinfo {volume} {89}},\ \bibinfo
  {pages} {226404} (\bibinfo {year} {2002})}\BibitemShut {NoStop}%
\bibitem [{\citenamefont {Egger}\ \emph {et~al.}(2003)\citenamefont {Egger},
  \citenamefont {Trauzettel}, \citenamefont {Chen},\ and\ \citenamefont
  {Siano}}]{Egger2003}%
  \BibitemOpen
  \bibfield  {author} {\bibinfo {author} {\bibfnamefont {R.}~\bibnamefont
  {Egger}}, \bibinfo {author} {\bibfnamefont {B.}~\bibnamefont {Trauzettel}},
  \bibinfo {author} {\bibfnamefont {S.}~\bibnamefont {Chen}}, \ and\ \bibinfo
  {author} {\bibfnamefont {F.}~\bibnamefont {Siano}},\ }\href@noop {}
  {\bibfield  {journal} {\bibinfo  {journal} {\njp}\ }\textbf {\bibinfo
  {volume} {5}},\ \bibinfo {pages} {117} (\bibinfo {year} {2003})}\BibitemShut
  {NoStop}%
\bibitem [{\citenamefont {Pham}\ \emph {et~al.}(2003)\citenamefont {Pham},
  \citenamefont {Pi\'echon}, \citenamefont {Imura},\ and\ \citenamefont
  {Lederer}}]{Pham2003}%
  \BibitemOpen
  \bibfield  {author} {\bibinfo {author} {\bibfnamefont {K.-V.}\ \bibnamefont
  {Pham}}, \bibinfo {author} {\bibfnamefont {F.}~\bibnamefont {Pi\'echon}},
  \bibinfo {author} {\bibfnamefont {K.-I.}\ \bibnamefont {Imura}}, \ and\
  \bibinfo {author} {\bibfnamefont {P.}~\bibnamefont {Lederer}},\ }\href
  {\doibase 10.1103/PhysRevB.68.205110} {\bibfield  {journal} {\bibinfo
  {journal} {Phys. Rev. B}\ }\textbf {\bibinfo {volume} {68}},\ \bibinfo
  {pages} {205110} (\bibinfo {year} {2003})}\BibitemShut {NoStop}%
\bibitem [{\citenamefont {Kazymyrenko}\ and\ \citenamefont
  {Doucot}(2005)}]{Kazymyrenko2005}%
  \BibitemOpen
  \bibfield  {author} {\bibinfo {author} {\bibfnamefont {K.}~\bibnamefont
  {Kazymyrenko}}\ and\ \bibinfo {author} {\bibfnamefont {B.}~\bibnamefont
  {Doucot}},\ }\href {\doibase 10.1103/PhysRevB.71.075110} {\bibfield
  {journal} {\bibinfo  {journal} {Phys. Rev. B}\ }\textbf {\bibinfo {volume}
  {71}},\ \bibinfo {pages} {075110} (\bibinfo {year} {2005})}\BibitemShut
  {NoStop}%
\bibitem [{\citenamefont {Das}\ and\ \citenamefont {Rao}(2008)}]{Das2008}%
  \BibitemOpen
  \bibfield  {author} {\bibinfo {author} {\bibfnamefont {S.}~\bibnamefont
  {Das}}\ and\ \bibinfo {author} {\bibfnamefont {S.}~\bibnamefont {Rao}},\
  }\href {\doibase 10.1103/PhysRevB.78.205421} {\bibfield  {journal} {\bibinfo
  {journal} {Phys. Rev. B}\ }\textbf {\bibinfo {volume} {78}},\ \bibinfo
  {pages} {205421} (\bibinfo {year} {2008})}\BibitemShut {NoStop}%
\bibitem [{\citenamefont {Bellazzini}\ \emph {et~al.}(2009)\citenamefont
  {Bellazzini}, \citenamefont {Mintchev},\ and\ \citenamefont
  {Sorba}}]{Bellazzini2009}%
  \BibitemOpen
  \bibfield  {author} {\bibinfo {author} {\bibfnamefont {B.}~\bibnamefont
  {Bellazzini}}, \bibinfo {author} {\bibfnamefont {M.}~\bibnamefont
  {Mintchev}}, \ and\ \bibinfo {author} {\bibfnamefont {P.}~\bibnamefont
  {Sorba}},\ }\href {\doibase 10.1103/PhysRevB.80.245441} {\bibfield  {journal}
  {\bibinfo  {journal} {Phys. Rev. B}\ }\textbf {\bibinfo {volume} {80}},\
  \bibinfo {pages} {245441} (\bibinfo {year} {2009})}\BibitemShut {NoStop}%
\bibitem [{\citenamefont {Aristov}\ \emph {et~al.}(2010)\citenamefont
  {Aristov}, \citenamefont {Dmitriev}, \citenamefont {Gornyi}, \citenamefont
  {Kachorovskii}, \citenamefont {Polyakov},\ and\ \citenamefont
  {Wolfle}}]{Aristov2010}%
  \BibitemOpen
  \bibfield  {author} {\bibinfo {author} {\bibfnamefont {D.~N.}\ \bibnamefont
  {Aristov}}, \bibinfo {author} {\bibfnamefont {A.~P.}\ \bibnamefont
  {Dmitriev}}, \bibinfo {author} {\bibfnamefont {I.~V.}\ \bibnamefont
  {Gornyi}}, \bibinfo {author} {\bibfnamefont {V.~Y.}\ \bibnamefont
  {Kachorovskii}}, \bibinfo {author} {\bibfnamefont {D.~G.}\ \bibnamefont
  {Polyakov}}, \ and\ \bibinfo {author} {\bibfnamefont {P.}~\bibnamefont
  {Wolfle}},\ }\href@noop {} {\bibfield  {journal} {\bibinfo  {journal} {\prl}\
  }\textbf {\bibinfo {volume} {105}},\ \bibinfo {pages} {266404} (\bibinfo
  {year} {2010})}\BibitemShut {NoStop}%
\bibitem [{\citenamefont {Aristov}(2011)}]{Aristov2011}%
  \BibitemOpen
  \bibfield  {author} {\bibinfo {author} {\bibfnamefont {D.~N.}\ \bibnamefont
  {Aristov}},\ }\href@noop {} {\bibfield  {journal} {\bibinfo  {journal}
  {\prb}\ }\textbf {\bibinfo {volume} {83}},\ \bibinfo {pages} {115446}
  (\bibinfo {year} {2011})}\BibitemShut {NoStop}%
\bibitem [{\citenamefont {Aristov}\ and\ \citenamefont
  {W\"olfle}(2011)}]{Aristov2011b}%
  \BibitemOpen
  \bibfield  {author} {\bibinfo {author} {\bibfnamefont {D.~N.}\ \bibnamefont
  {Aristov}}\ and\ \bibinfo {author} {\bibfnamefont {P.}~\bibnamefont
  {W\"olfle}},\ }\href {\doibase 10.1103/PhysRevB.84.155426} {\bibfield
  {journal} {\bibinfo  {journal} {Phys. Rev. B}\ }\textbf {\bibinfo {volume}
  {84}},\ \bibinfo {pages} {155426} (\bibinfo {year} {2011})}\BibitemShut
  {NoStop}%
\bibitem [{\citenamefont {Hou}\ and\ \citenamefont {Chamon}(2008)}]{Hou2008}%
  \BibitemOpen
  \bibfield  {author} {\bibinfo {author} {\bibfnamefont {C.-Y.}\ \bibnamefont
  {Hou}}\ and\ \bibinfo {author} {\bibfnamefont {C.}~\bibnamefont {Chamon}},\
  }\href@noop {} {\bibfield  {journal} {\bibinfo  {journal} {\prb}\ }\textbf
  {\bibinfo {volume} {77}},\ \bibinfo {pages} {155422} (\bibinfo {year}
  {2008})}\BibitemShut {NoStop}%
\bibitem [{\citenamefont {Safi}\ and\ \citenamefont {Schulz}(1995)}]{Safi1995}%
  \BibitemOpen
  \bibfield  {author} {\bibinfo {author} {\bibfnamefont {I.}~\bibnamefont
  {Safi}}\ and\ \bibinfo {author} {\bibfnamefont {H.~J.}\ \bibnamefont
  {Schulz}},\ }\href {\doibase 10.1103/PhysRevB.52.R17040} {\bibfield
  {journal} {\bibinfo  {journal} {Phys. Rev. B}\ }\textbf {\bibinfo {volume}
  {52}},\ \bibinfo {pages} {R17040} (\bibinfo {year} {1995})}\BibitemShut
  {NoStop}%
\bibitem [{\citenamefont {Maslov}\ and\ \citenamefont
  {Stone}(1995)}]{Maslov1995}%
  \BibitemOpen
  \bibfield  {author} {\bibinfo {author} {\bibfnamefont {D.~L.}\ \bibnamefont
  {Maslov}}\ and\ \bibinfo {author} {\bibfnamefont {M.}~\bibnamefont {Stone}},\
  }\href {\doibase 10.1103/PhysRevB.52.R5539} {\bibfield  {journal} {\bibinfo
  {journal} {Phys. Rev. B}\ }\textbf {\bibinfo {volume} {52}},\ \bibinfo
  {pages} {R5539} (\bibinfo {year} {1995})}\BibitemShut {NoStop}%
\bibitem [{\citenamefont {Aristov}\ and\ \citenamefont
  {W\"olfle}(2012)}]{Aristov2012}%
  \BibitemOpen
  \bibfield  {author} {\bibinfo {author} {\bibfnamefont {D.~N.}\ \bibnamefont
  {Aristov}}\ and\ \bibinfo {author} {\bibfnamefont {P.}~\bibnamefont
  {W\"olfle}},\ }\href {\doibase 10.1103/PhysRevB.86.035137} {\bibfield
  {journal} {\bibinfo  {journal} {Phys. Rev. B}\ }\textbf {\bibinfo {volume}
  {86}},\ \bibinfo {pages} {035137} (\bibinfo {year} {2012})}\BibitemShut
  {NoStop}%
\bibitem [{\citenamefont {Hou}\ \emph {et~al.}(2012)\citenamefont {Hou},
  \citenamefont {Rahmani}, \citenamefont {Feiguin},\ and\ \citenamefont
  {Chamon}}]{Hou2012}%
  \BibitemOpen
  \bibfield  {author} {\bibinfo {author} {\bibfnamefont {C.-Y.}\ \bibnamefont
  {Hou}}, \bibinfo {author} {\bibfnamefont {A.}~\bibnamefont {Rahmani}},
  \bibinfo {author} {\bibfnamefont {A.~E.}\ \bibnamefont {Feiguin}}, \ and\
  \bibinfo {author} {\bibfnamefont {C.}~\bibnamefont {Chamon}},\ }\href
  {\doibase 10.1103/PhysRevB.86.075451} {\bibfield  {journal} {\bibinfo
  {journal} {Phys. Rev. B}\ }\textbf {\bibinfo {volume} {86}},\ \bibinfo
  {pages} {075451} (\bibinfo {year} {2012})}\BibitemShut {NoStop}%
\bibitem [{\citenamefont {Aristov}\ and\ \citenamefont
  {W\"olfle}(2013)}]{Aristov2013}%
  \BibitemOpen
  \bibfield  {author} {\bibinfo {author} {\bibfnamefont {D.~N.}\ \bibnamefont
  {Aristov}}\ and\ \bibinfo {author} {\bibfnamefont {P.}~\bibnamefont
  {W\"olfle}},\ }\href {\doibase 10.1103/PhysRevB.88.075131} {\bibfield
  {journal} {\bibinfo  {journal} {Phys. Rev. B}\ }\textbf {\bibinfo {volume}
  {88}},\ \bibinfo {pages} {075131} (\bibinfo {year} {2013})}\BibitemShut
  {NoStop}%
\bibitem [{\citenamefont {Cirillo}\ \emph {et~al.}(2011)\citenamefont
  {Cirillo}, \citenamefont {Mancini}, \citenamefont {Giuliano},\ and\
  \citenamefont {Sodano}}]{Cirillo:2011uq}%
  \BibitemOpen
  \bibfield  {author} {\bibinfo {author} {\bibfnamefont {A.}~\bibnamefont
  {Cirillo}}, \bibinfo {author} {\bibfnamefont {M.}~\bibnamefont {Mancini}},
  \bibinfo {author} {\bibfnamefont {D.}~\bibnamefont {Giuliano}}, \ and\
  \bibinfo {author} {\bibfnamefont {P.}~\bibnamefont {Sodano}},\ }\href
  {\doibase http://dx.doi.org/10.1016/j.nuclphysb.2011.06.014} {\bibfield
  {journal} {\bibinfo  {journal} {Nuclear Physics B}\ }\textbf {\bibinfo
  {volume} {852}},\ \bibinfo {pages} {235 } (\bibinfo {year}
  {2011})}\BibitemShut {NoStop}%
\bibitem [{\citenamefont {Giuliano}\ and\ \citenamefont
  {Sodano}(2013)}]{Giuliano:2013fk}%
  \BibitemOpen
  \bibfield  {author} {\bibinfo {author} {\bibfnamefont {D.}~\bibnamefont
  {Giuliano}}\ and\ \bibinfo {author} {\bibfnamefont {P.}~\bibnamefont
  {Sodano}},\ }\href {http://stacks.iop.org/0295-5075/103/i=5/a=57006}
  {\bibfield  {journal} {\bibinfo  {journal} {EPL (Europhysics Letters)}\
  }\textbf {\bibinfo {volume} {103}},\ \bibinfo {pages} {57006} (\bibinfo
  {year} {2013})}\BibitemShut {NoStop}%
\end{thebibliography}%

\end{document}